\documentclass[journal]{vgtc}  

\newcommand{\comment}[1]{}

\newcommand{\imc}{Imaging Mass Cytometry}
\newcommand{\histo}{histoCAT}
\newcommand{\mimc}{ImaCytE}

\newcommand{\omi}{{omics}}
\newcommand{\mytitle}{Visual cohort comparison for spatial single-cell \omi{}-data }

\newenvironment{myfont}{\fontfamily{qcr}\selectfont}{\par}
\DeclareTextFontCommand{\textmyfont}{\myfont}

\ifpdf
  \pdfoutput=1\relax                   
  \usepackage{graphicx}                
  \DeclareGraphicsExtensions{.pdf,.png,.jpg,.jpeg} 
\else
  \usepackage{graphicx}                
  \DeclareGraphicsExtensions{.eps}     
\fi%

\graphicspath{{figures/}{pictures/}{images/}{./}} 

\usepackage{microtype}                 
\PassOptionsToPackage{warn}{textcomp}  
\usepackage{textcomp}                  
\usepackage{mathptmx}                  
\usepackage{times}                     
\usepackage{cite}                      
\usepackage{tabu}                      
\usepackage{booktabs}                  
\usepackage{tikz}
\usepackage{colortbl}
\usepackage{enumitem}
\usepackage{multirow}

\RequirePackage{etoolbox}
\pretocmd\PackageWarning{%
    \edef\pkgname{#1}\edef\hyperrefname{hyperref}%
    \ifx\pkgname\hyperrefname
        \expandafter\gobblethree
    \fi
}{}{\undefined}
\newcommand*{\gobblethree}[3]{}

\definecolor{CBpurple}{HTML}{bc80bd}
\definecolor{CBpink}{HTML}{fccde5}
\definecolor{CBgreen}{HTML}{b3de69}
\definecolor{CByellow}{HTML}{ffed6f}
\definecolor{CBturqoise}{HTML}{8dd3c7}
\definecolor{CBgrey}{HTML}{d9d9d9}

\newcommand{\envtype}[1]{%
    \begin{tikzpicture}
        \draw[#1, fill=#1] (0,0) circle (3pt);
    \end{tikzpicture}%
}

\newcommand{\microenv}[2]{%
    \envtype{#1}~%
    \begin{tikzpicture}
        \draw (0,0) -- (0,6pt);
    \end{tikzpicture}~%
    \envtype{#2}%
}

\newcommand{\microempty}{%
    \begin{tikzpicture}
        \draw(0,0) circle (3pt);
        \draw (-3pt,-3pt) -- (3pt,3pt);
    \end{tikzpicture}%
}

\newcommand{\qp}{%
    \begin{tikzpicture}
        \draw[CBturqoise, fill=CBturqoise] (0,0) circle (2pt);
    \end{tikzpicture}%
}

\newcommand{\qd}{\qp\,\qp\,\qp}
\newcommand{\qz}{\qp\,\qp}

\newcommand{\neutral}{%
    \begin{tikzpicture}
        \draw (0,0) circle (1.8pt);
    \end{tikzpicture}%
}

\newcommand{\qq}[1]{%
``\textit{#1}''%
}

\onlineid{0}

\vgtccategory{Research}
\vgtcpapertype{application/design study}

\title{\mytitle}

\author{Antonios Somarakis,
        Marieke E. Ijsselsteijn,
         Sietse J. Luk,
        Boyd Kenkhuis,\\
        Noel F.C.C. de Miranda,
        Boudewijn P.F. Lelieveldt, 
        and Thomas H\"ollt}
\authorfooter{
\item
 Antonios Somarakis and Boudewijn P.F. Lelieveldt are with the Division of Image Processing, Department of Radiology, Leiden University Medical Center, the Netherlands.
\item
 Marieke E. Ijsselsteijn and Noel F.C.C. de Miranda are with the Immunogenomics group, Department of Pathology, Leiden University Medical Center, the Netherlands.
\item
 Sietse J. Luk is with the Hematology Department, Leiden University Medical Center, the Netherlands.
\item
 Boyd Kenkhuis is with the Human Genetics Departments, Leiden University Medical Center, the Netherlands.
\item
 Thomas H\"ollt is with the Computer Graphics and Visualization Group, TU Delft and the Leiden Computational Biology Center, Leiden University Medical Center, The Netherlands. E-mail: T.Hollt-1@tudelft.nl
}

\shortauthortitle{Somarakis \MakeLowercase{\textit{et al.}}: \mytitle}

\abstract{%
Spatially-resolved \omi{}-data enable researchers to precisely distinguish cell types in tissue and explore their spatial interactions, enabling deep understanding of tissue functionality.
To understand what causes or deteriorates a disease and identify related biomarkers, clinical researchers regularly perform large-scale cohort studies, requiring the comparison of such data at cellular level.
In such studies, with little \textit{a-priori} knowledge of what to expect in the data, explorative data analysis is a necessity.
Here, we present an interactive visual analysis workflow for the comparison of cohorts of spatially-resolved \omi{}-data.
Our workflow allows the comparative analysis of two cohorts based on multiple levels-of-detail, from
simple abundance of contained cell types over complex co-localization patterns to individual comparison of complete tissue images.
As a result, the workflow enables the identification of cohort-differentiating features, as well as outlier samples at any stage of the workflow.
During the development of the workflow, we continuously consulted with domain experts.
To show the effectiveness of the workflow, we conducted multiple case studies with domain experts from different application areas and with different data modalities.}

\keywords{Visual analytics, Imaging Mass Cytometry, Vectra, spatially-resolved data, single-cell \omi{}-data, Visual comparison}

\CCScatlist{ 
 \CCScat{K.6.1}{Management of Computing and Information Systems}%
{Project and People Management}{Life Cycle};
 \CCScat{K.7.m}{The Computing Profession}{Miscellaneous}{Ethics}
}

\teaser{
  \centering
  \includegraphics[width=0.9\linewidth]{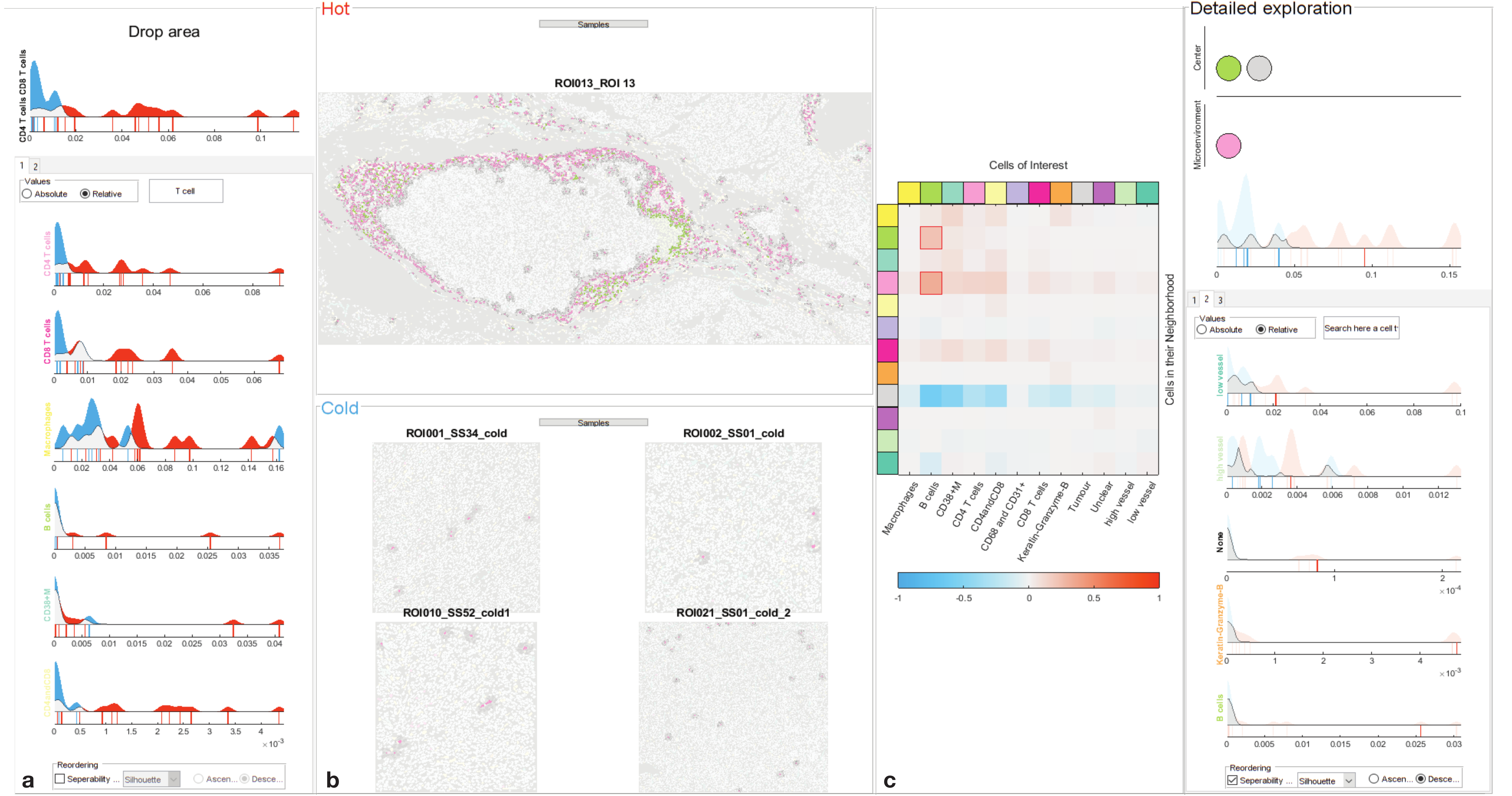}
  \caption{\textbf{Screenshot of our integrated system} including the view for the comparison based on the cell abundance using raincloud plots (a), the tissue view, showing selected samples of the two cohorts (b), and the multi-cellular microenvironment comparison view using a difference heatmap and raincloud plots (c).}
	\label{fig:System}
}

\vgtcinsertpkg

\begin{document}



\maketitle

\section{Introduction}\label{introduction}
%
Omics-data describe biochemical properties, such as genomics, transcriptomics, proteomics, or metabolomics of biological systems~\cite{conesa2019making}, such as cells.
In recent years, high-resolution spatial measurements of such systems have become available.
State of the art spatially-resolved \omi\ modalities \cite{Ke2013,Giesen2014,Crosetto2015,Lee2015,Goltsev2018} enable the precise characterization of cellular populations in tissue, enabling the discovery and identification of novel cell types\cite{VanUnen2017} in large cohorts of samples.
Information about the cell type, in combination with the specific location of each cell creates many heterogeneous multi-cellular patterns.

With the identification of these multi-cellular patterns, a crucial question arises; are such patterns correlated with clinical information, such as survival rate?
Current research findings~\cite{Ali2020,Keren2020,Jackson2020} support the clinical importance of analysing spatial multi-cellular interactions.
Hence, the development of workflows for the systematic comparison of cohorts consisting of spatially-resolved \omi{}-data with specific clinical characteristics is essential for the understanding of tissue functionality.

In the majority of life-science studies, the comparison of cohorts of samples is based on statistical comparison of predefined finite number of elements~\cite{newschaffer2000causes,Yuan2012,Nagaishi2011,Robert2014}.
However, traditional statistical approaches, based on prior knowledge pose the risk of missing unexpected correlations and cannot capture the vast combinatorial space~\cite{Cibulski2016} of spatial configurations for all different cell types.
Moreover, they depend on high quality input which often cannot be guaranteed with single-cell omics-data due to uncertainty in cell segmentation and cell type identification.
Comparative visualization~\cite{pagendarm1995comparative} can provide useful insights into the differentiating factors of two cohorts and enables the interactive, data-driven exploration of the vast combinatorial space while simultaneously investigating the biological relevance and plausibility of findings with regard to the preprocessing.
%
%

Here, we extended our previous work focused on the identification and exploration of multi-cellular spatial interactions in single-cell omics-data~\cite{Somarakis2019} to enable interactive comparison of cohorts of such data. 
%
The main goals are to identify the characteristics that differentiate a cohort, explore the cohorts' heterogeneity and relate these characteristics directly to the tissue.
In some cases, just the comparison of the cell types abundance is adequate to differentiate cohorts.
In other cases, a detailed comparison of contained cells and their specific neighborhoods, i.e. microenvironments is needed.

We propose an interactive, data-driven cohort comparison workflow.
More specifically the main contributions of this paper are:
\begin{enumerate}[leftmargin=6mm]
    \item A workflow for the comparison of cohorts of spatially-resolved single-cell \omi{}-data, specifically addressing the following tasks
    \begin{itemize}[leftmargin=5mm]
        \item[\textbf{T1}] compare cohorts based on the abundance of different cell types,
        \item[\textbf{T2}] compare cohorts based on multi-cellular microenvironments,
        \item[\textbf{T3}] detect outliers within each cohort, and
        \item[\textbf{T4}] relate findings to their spatial position.
    \end{itemize}
    \item A protoype implementation of the described workflow 
\end{enumerate}

The remainder of this paper is structured as follows.
We present related work in \autoref{related_work}, followed by a brief description of target users, input data and tasks in \autoref{abstraction}.
In \autoref{Workflow}, we describe the rationale behind our visual design and implementation in our prototype.
We present a set of case studies and user feedback in \autoref{case_study}. 
Finally, we discuss the limitations of our work and conclude in \autoref{conclusion}.

\section{Related Work}\label{related_work}
The visual analytics community spent considerable effort on approaches for the exploration of cohorts of medical data combining spatial and non-spatial features.
Preim et al.~\cite{Preim2016} provide an overview of image-centric approaches~\cite{Dzyubachyk2013,Steenwijk2010,Zhang} focused on the exploration of large imaging cohorts and derived attributes.
For the data analysis, these approaches share linking of attribute views with image views to provide context, visual queries for direct feedback, and interactive definition of groups of attributes.
They typically deal with traditional medical imaging databases, such as those acquired by computed tomography (CT) or magnet resonance imaging (MRI).

Dealing with microscopic images, Screenit~\cite{dinkla2017screenit} offers a system of linked views, similar to our system, to explore the drug screening results of cell cultures at multiple levels of detail. 
However, only recently, spatially-resolved \omi{}-data~\cite{Ke2013, Crosetto2015, Giesen2014} have become a standard tool for the exploration of tissue structure at the cellular level.
Consequently, only few visual analysis tools exist that address the specific needs of such data.
Facetto~\cite{Krueger2020} is a scalable framework that allows hierarchical cell type identification in large multiplexed images.
\histo~\cite{Schapiro2017} enables the identification of cell types and the significant pairwise spatial interactions between them.
CytoMAP~\cite{stoltzfus2020cytomap} offers an extensive toolbox for the exploration of tissue structure based on the analysis of spatial interactions.
In our previous work on \mimc~\cite{Somarakis2019}, we propose an interactive exploratory pipeline for cell type identification and neighborhood analysis in spatial single-cell data.
Minerva~\cite{Rashid2020.03.27.001834} extends such exploration concepts with storytelling tools, to support communication and sharing of results.
All of the above focus on the identification or exploration of cell types or significant multi-cellular interactions in a single cohort of spatial single-cell data.
Here, we use some of the concepts introduced in these works and extend them to introduce the first workflow for comparative analysis of two cohorts of such data, based on the abundance of cell types, as well as  colocation patterns.

Based on a survey on existing comparative visualization tools~\cite{Gleicher2011}, Gleicher et al. define a taxonomy that divides comparative visualization into juxtaposition (side-by-side placement), superposition (layering), and explicit encoding.
A large body of work on comparative visualization for individual images exist.
For example, Blaas et al.~\cite{Dzyubachyk2013} combine superposition with explicit coding of the differences using complementary colors for the comparands, which cancels out in regions without differences. We use the same technique in some of our charts.
Lindemann et al.~\cite{Bronstein}, Maries et al.~\cite{Maries2013} and Ma et al.~\cite{Ma2017} utilize juxtaposition in an interactive comparative visualization pipeline for one-to-one comparison of segmentation results of brain imaging data. Juxtaposition for the comparison of images is also utilized in our work.

 Schmidt et al.~\cite{Schmidt2013} facilitate the comparison of images with small differences within an ensemble.
Raidou et al.~\cite{Raidou2018} compare volume data and corresponding segmentations of bladders to explore the results of longitudinal radiotherapy treatment studies. 
Both works focus on all-to-all comparison of (3D) images in a single group, compared to the between-cohort comparison presented in this work.
%
Basole et al.~\cite{Basole2015} as well as Wagner et al.~\cite{Wagner2019} propose  pipelines for the comparison of two cohorts.
In their comparison workflow they use the same visual enocdings in order to compare the cohorts as a whole and simultaneously provide information for the intra-cohort heterogeneity, similar to the visual encodings we utilize in our system. Both approaches are limited to non-spatial healthcare data, though.
Zhang et al.~\cite{zhang2017comparative} present a visual analytics approach to compare two cohorts of diffusion tensor images.
While we took some inspiration from their work, such as using complementary colors for the two cohorts that cancel each other out when overlapping, ultimately, the solutions described in their work are specific to tensor data and not easily transferrable to the spatial single-cell data described here.
%
%
%
%
%

\section{Abstraction}\label{abstraction}
Recent developments in the spatially-resolved \omi\ field manifest a wide variety of available modalities~\cite{Lee2015,Femino1998,Giesen2014,Keren2019}.
These technologies measure transcriptomics or proteomics information at sub-cellular resolution, resulting in high-resolution image data with tens to thousands of values per pixel.
Since researchers are interested in this information per cell, rather than per pixel, these images are typically pre-processed by segmenting individual cells and aggregating the values of the segmented pixels.
Based on this aggregated information and potentially further features like morphology, the function and type of the segmented cells can be identified~\cite{Schapiro2017}.
Both, cell segmentation~\cite{schuffler2015automatic, Schapiro2017}, as well as cell type identification~\cite{Krueger2020, Schapiro2017, stoltzfus2020cytomap, Somarakis2019} in this kind of data is an active research topic.
Large variations in cellular morphology and different quality of marker staining, among others, can lead to a considerable amount of uncertainty in the result of these preprocessing steps, making the validation, for example by referencing the actual images, during comparison imperative.


\subsection{Target Users and Goals}\label{workflow_users}

Our proposed workflow is targeted at clinical researchers who want to analyze their own data, for example to do an initial exploration of the data to form hypotheses.
Typical goals when doing comparative analysis of two cohorts of spatial single-cell data could be the identification of cell types that are abundant in one cohort but not the other or cell co-localization patterns that are correlated with one of the cohorts.
Such correlations or biomarkers~\cite{Mayeux2004} can be used for prognosis, monitoring or therapy of disease.
While scripting in python or R is becoming more common in the domain, all our collaborators prefer visual exploration through GUI interfaces.
Our proposed workflow is the first such visual exploration system that supports the comparative analysis of two cohorts of spatial single-cell data.

\subsection{Input Data}\label{input_data}

The overarching goal of our workflow is the comparison of two cohorts of spatially-resolved \omi\ data as briefly introduced above.
A single cohort consists of a set of samples, i.e., segmented and classified images as described above.
Depending on the goal of the study, the samples consist of multiple images from a single subject or an arbitrary number of samples from multiple subjects.
Typically, the two cohorts describe different populations, for instance, cancer patients who respond well to treatment in one cohort and those who respond worse in the second.
A typical cohort consists of tens to hundreds of images, each consisting of thousands of segmented cells.

In a typical study, tens to hundreds of different cell types will be identified.
The granularity depends on the goal of the study, as well as the data modality.
For example, the Vectra imaging system~\cite{Nghiem2016} measures only a few different proteins (i.e. 4 in the case study in \autoref{study_alzheimers}).
Assuming differentiation into only low and high abundance, this results in an upper limit of $2^4 = 16$ differentiable cell types.
Other systems, such as \imc{}, allow the measurement of up to 40 proteins, such that the number of cell types is limited rather by which types are of interest for the given study.
A broad study would capture in the order of a hundred different cell types.

For each sample, we store the segmentation mask including a cell type label, i.e. class, for each segmented cell.
Based on the cell segmentation mask, we derive the microenvironment for each cell.
The microenvironment consists of the cell types and their abundance in the neighborhood of the given cell.
We store the corresponding information per cell as a list of all cells that are contained in the microenvironment.
The microenvironment of a cell varies according to the resolution of the modality and the type of sample.
For example, in a tumor crowded with compact cells we would consider cells belonging to the microenvironment in a smaller distance, compared to brain tissue, where interacting cells can be further apart.
Therefore, the distance defining the microenvironment of a cell is specified by the user.
Typically, the microenvironment of a cell consists of no more than some tens of cells.

\subsection{Identified Tasks}\label{id_tasks}


In the following, we describe a set of tasks that we have identified in close collaboration with our domain expert partners from the pathology department at LUMC (co-authors of this manuscript).
In general, we compare the two cohorts, based on the contained samples.
The first step of the workflow is comparing the cohorts according to the abundance of different cell types per sample~(\textbf{T1}).
This allows a simple differentiation of the cohorts based on the contained cells.
In the second step, we further want to identify patterns in the cells' microenvironments that differentiate the cohorts.
In~\textbf{T2}, we compare cohorts based on multi-cellular microenvironments.
Throughout the process we support visual detection of outliers within each cohort~(\textbf{T3}), according to the abundance of contained cells and their microenvironments, and relate any findings to their spatial position~(\textbf{T4}).

In the following, we describe and abstract \textbf{T1}-\textbf{T4} in more detail using Brehmer and Munzners task typology~\cite{Brehmer2013}.
For references to this typology, we use a \textmyfont{mono-spaced} font.

\begin{itemize}
        \item
            [\textbf{T1}] \textbf{Cohort comparison  based on the abundance of different cell types and combinations thereof in cohort samples.}
            The relative abundance of a cell type in the samples forming a cohort and how much a specific subject deviates from the distribution within the cohort are important clinical biomarkers.
            As cell types can be of different granularity, it should also be possible to compare the cohorts, based on combinations of cell types.
            A trivial example is differentiating a cohort of cancer patients and a cohort of healthy subjects by comparing the abundance of tumor cells in the contained samples, where ``tumor cells'' can be a single cell type, or a combination of cell types according to a more fine grained definition.
            In this task \textbf{T1}, the user \textmyfont{compares} the two cohorts based on the abundance of different cell types within samples forming the cohort \textmyfont{discovering} and \textmyfont{locating} the cell type(s) that differentiate the two cohorts.
            The input for \textbf{T1} is the abundance of each cell type for each sample that we \textmyfont{summarize} as distributions over all samples in one cohort.
            The output is a list of cell types that differentiate the two cohorts.
        \item
            [\textbf{T2}]\textbf{Cohort comparison based on multi-cellular microenvironments.}
            The goal of \textbf{T2} is to compare the two cohorts according to the spatial co-localization patterns of each sample, as the comparison only based on cell type abundance is not enough to assess tissue functionality.
            Domain researchers hypothesize that tissue functionality also depends on the cell's interactions with other cells.
            While co-localization does not automatically lead to such interactions, it is a pre-condition.
            We facilitate the identification of such spatial features by breaking this task down into a high-level comparison, based on how often any two cell types are spatially co-located (\textbf{T2.a}), and a detail comparison where complex user-defined microenvironments can be explored (\textbf{T2.b}).
            In task \textbf{T2.a}, the user \textmyfont{discovers} combinations of two cell types that are most differentiating between the two cohorts.
            The input for this task is the abundance of each combination of two cell types in a microenvironment within the cohort sample.
            The output is a combination of two cell types to be used for further \textmyfont{exploration}.
            In task \textbf{T2.b}, the user further \textmyfont{explores} and \textmyfont{compares} the two cohorts based on more complex microenvironment compositions.
            Therefore, the user \textmyfont{produces} these more complex microenvironments by combining different cell types, typically starting with the combination found in \textbf{T2.a}.
            The input for \textbf{T2.b} is the complete set of cell microenvironments, optionally filtered to those including the combination of interest \textmyfont{discovered} in \textbf{T2.a}.
            The output is a set of detailed microenvironments differentiating the two cohorts.
        \item
            [\textbf{T3}]\textbf{Outlier detection within each cohort.}
            Detecting outliers within a cohort can provide additional important clinical information.
            For example subjects with different stages of a disease in the same cohort might exhibit different cell profiles~\cite{VanUnen2016}.
            Therefore, \textbf{T3} consists of \textmyfont{identifying} and \textmyfont{locating} outlying samples and their corresponding features identified in \textbf{T1} and \textbf{T2}.
            The input to this task is the abundance of cells and their microenvironments, as \textmyfont{identified} in \textbf{T1} and \textbf{T2}.
            The output is a list of outlying samples.
        \item
            [\textbf{T4}]\textbf{Relate findings to their spatial position.}
            As described above, \textbf{T1}-\textbf{T3} can be carried out based on cell abundance and microenvironment descriptions per sample, without consulting the actual imaging data.
            However, to verify individual findings we inspect the cells and their neighborhoods in their tissue context. 
            Therefore, \textbf{T4} relates any findings to their spatial position. 
            The analyst \textmyfont{locates} the structure of interest in their spatial location and  \textmyfont{identifies} issues that were not apparent in the abstract representation.
            The input to \textbf{T4} are the segmented images and a structure of interest found with \textbf{T1}-\textbf{T3}, the output is a verified or rejected finding from \textbf{T1}-\textbf{T3}.
\end{itemize}

\comment{\subsection{\textcolor{red}{Alternative task abstraction}}

We developed our workflow in order to support four main tasks;
    \begin{itemize}
        \item[\textbf{T1}] cohorts comparison  based on the abundance of different cell types,
        \item[\textbf{T2}] cohorts comparison based on multi-cellular spatial interactions,
        \item[\textbf{T3}] outliers detection  in each cohort, and
        \item[\textbf{T4}] relate findings to their spatial position
    \end{itemize}
, as they are described in \autoref{introduction}.

\autoref{fig:Tasks} illustrates the order the tasks are performed in the workflow.
The workflow starts with the comparison among two cohorts based on the abundance of the different cell types (\textbf{T1}) and continues with a more targeted comparison based on the plethora of multi-cellular spatial interactions (\textbf{T2}). 
During both, T1 and T2 the user can detect outliers in the cohorts (\textbf{T3}) and also link any findings to their spatial position (\textbf{T4}).

We describe in a more detail and abstract way the aforementioned tasks using the the Brehmer and Munzners task typology~\cite{Brehmer2013}.
For the references to this typology  we use a \textmyfont{mono-spaced} font.

The user in task \textbf{T1} \textmyfont{compares} the distributions of the cohorts to \textmyfont{discover} and \textmyfont{locate} the cell type(s) that better differentiate the cohorts
Input for task \textbf{T1} is the cell type abundance of each sample of a cohort. The output is for each cell type and cohort the distribution of each samples abundance.

In an average study, there are 30 different cell types and  5 cells in each microenvironment resulting in a a couple of millions spatial interactions. 
Therefore, we break task \textbf{T2} in two subtasks to guide the user during the visual comparison of the spatial interactions. 
In task \textbf{T2.a} the user from a \textmyfont{summary} of the spatial interactions for every two cell types, \textmyfont{locates} and\textmyfont{discovers} those differentiating the most the two cohorts.
Input of this task is the cell type and location of each cell and output the amount each pairwise interaction differentiates the two cohorts.
Afterwards, in task \textbf{T2.b} the user further \textmyfont{compares} the cohorts \textmyfont{exploring} the spatial interactions of multiples cell types.
The input in \textbf{T2.b} is the same as in \textbf{T2.a}.
The output now is the amount of cells each sample has for every possible cell type combination.

The heterogeneity exploration in a cohort is one of the main goals of the user in this study.
Hence, in task \textbf{T3} the user \textmyfont{locates} the samples and \textmyfont{identifies} the outliers during \textbf{T1} and \textbf{T2} tasks.
Input is the amount of cells  that fulfil either a specific cell type or a spatial interaction among multiple cell types.
Output is their relative position regarding the distribution of their cohort.

The example of floating cancer cells, described in detail in our previous study~\cite{somarakis2019imacyte} underlines the importance of the relation of a finding to its spatial location.
Also, in this study in task \textbf{T4} the user \textmyfont{verifies} any finding \textmyfont{locating} it in its spatial location.
Input is the the spatial location of all cells and output the location of a finding.

}

\section{Workflow}\label{Workflow}
%
We designed a workflow to support the four tasks, identified and described in \autoref{id_tasks} and implemented it in a multiple-linked-views system, shown in \autoref{fig:System}.
The system is divided in three main blocks, where the left (\autoref{fig:System}a) and right (\autoref{fig:System}c) blocks support \textbf{T1} and \textbf{T2}, respectively by comparing the cohorts based on their cell type abundance and  spatial interactions.
\textbf{T4} relies on the inspection of tissue samples and supports \textbf{T1}-\textbf{T3}. Therefore, we show the corresponding images between the views (\autoref{fig:System}b) for \textbf{T1} and \textbf{T2} to support the user in directly making the connection for structures identified in any of the tasks to their spatial position.
All views allow filtering the data to support visual outlier detection (\textbf{T3}).


\subsection {Comparison Based on Cell Type Abundance}\label{non-spatial}

\begin{figure}[t!]
 \centering
 \includegraphics[width=0.9\linewidth]{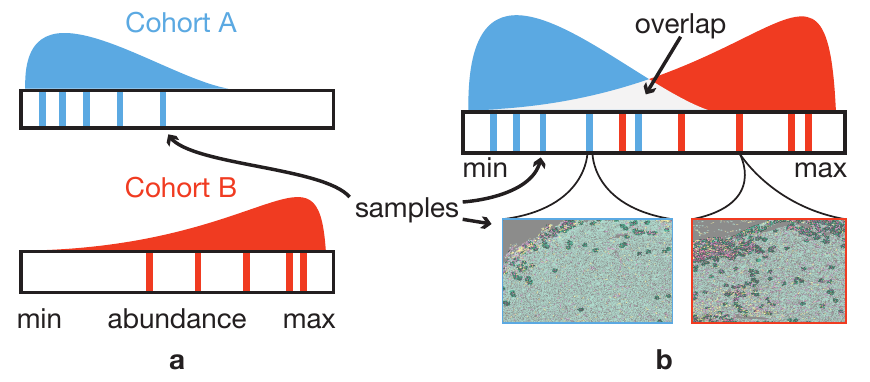}
 \vspace{-3mm}
 \caption{\textbf{Comparison of two cohorts based on a cell type abundance.}
(a) Individual raincloud plots for two cohorts showing the distribution (cloud) of samples (rain drops) according to the abundance of a contained cell type.
 (b) Superposition makes the difference visible by the large amount of color and small light-gray overlap area in the area chart.
 \vspace{-5mm}
}
 \label{fig:one_pdf}
\end{figure}

\begin{figure}[b!]
 \centering
 \includegraphics[width=0.9\linewidth]{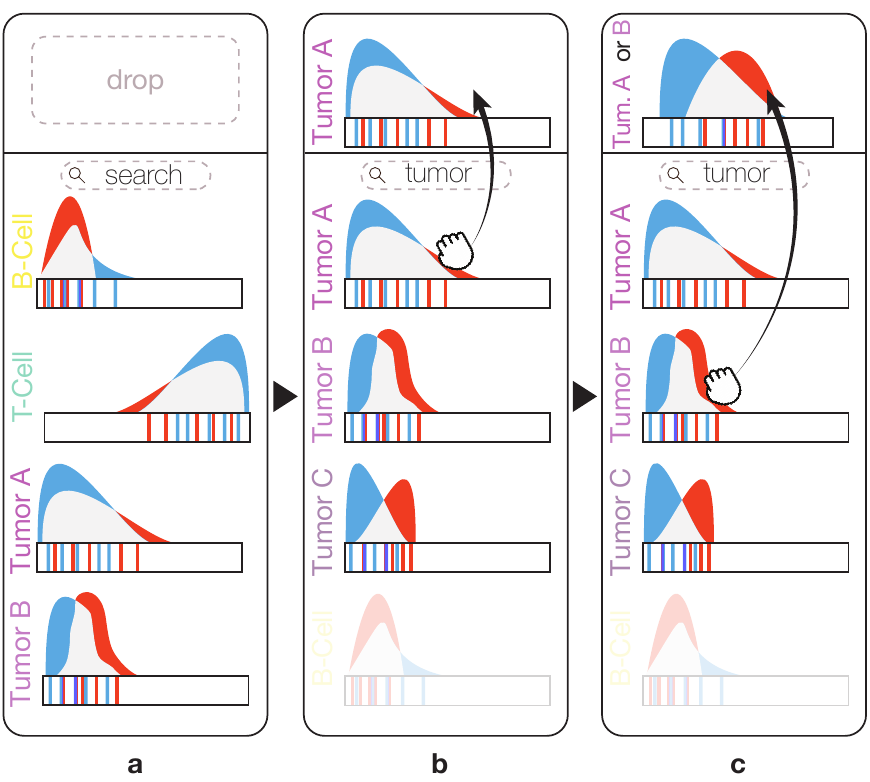}
 \vspace{-3mm}
 \caption{\textbf{Exploration using the raincloud plots.} Searching for ``tumor'' reorders the raincloud plots by placing the plots corresponding to cell types containing the term `tumor'' in their label on top of the list (b).
 Dragging a raincloud plot and dropping it in the drop area (b,c), creates a new raincloud plot depicting the abundance of the cell types represented from the accumulated dropped raincloud plots.
}
 \label{fig:drag_n_drop}
\end{figure}

In the first step, we are interested to compare two cohorts according to the abundance of the different existing cell types in each of the contained samples (\textbf{T1}) and visually detect possible outliers in each of the cohorts (\textbf{T3}).
Therefore, we first compute the number of cells of each type within each sample and then visualize the distribution of samples within both cohorts according to this value by superposing two simplified versions of raincloud plots~\cite{Allen2018}.
This plot consists of
a density (estimated using a kernel density estimate) plot showing the distribution of samples (the \emph{cloud}) above a one-dimensional scatterplot with vertical lines as marks for the individual samples (\emph{rain}-drops).
This combination has proven very effective for our goals in \textbf{T1}-\textbf{T3}.
The superposition of the density plots has shown to be very effective for the comparison of two distributions~\cite{v-plots2020}.
Both, the density plot~\cite{correll2018} and the one-dimensional scatterplot~\cite{kampstra2008beanplot}, support visual detection of outliers.
Furthermore, individual samples can be efficiently selected in the scatterplot for filtering.
%
Additionally, for easier comparison between samples of different sizes, we enable the user to select whether the x-axis should represent the number of cells either as absolute values, or relative to the number of cells in that sample.
As our primary goal is the comparison of the two cohorts, rather than the shape of individual plots, we want to emphasize the differences, rather than the commonalities.
Therefore, following the same principle as Blaas et. al.~\cite{Dzyubachyk2013}, we use complementary colors for the two cohorts, i.e. blue and orange and blend the PDFs additively to receive a neutral light-gray in the overlapping areas as shown in~\autoref{fig:one_pdf}b.
The resulting raincloud plot allows the comparison of the composition of the the two cohorts, according to the abundance of a single cell type within the contained samples.
To allow the inspection of these distributions for all cell types, we use a small multiples approach~\cite[Chapter~4]{tufte1990envisioning} and show the raincloud plots for several cell types in the same view (\autoref{fig:drag_n_drop}).

As indicated in \autoref{input_data}, some studies can contain up to $100$ different cell types.
Finding a specific type of interest or the types that are the most differentiating for the two cohorts manually is not feasible in such a case.
Therefore, we provide the possibility to sort the plots according to how well the corresponding distributions of the two cohorts separate, by default using the Silhouette metric~\cite{Rousseeuw1987}, as it is invariant to the range of the input data.
For advanced users we provide a set of other metrics, such as Dunn's index~\cite{Bezdek1995} which is efficient for compact and well separated clusters.
In addition, we provide filtering by means of a textual search box (\autoref{fig:drag_n_drop}a), based on the cell labels in the input data.
Typing, for example \textit{tumor} in this box will bring plots
with the term \textit{tumor} in their provided label to the top of the view (\autoref{fig:drag_n_drop}b).



In some cases, the analyst might also be interested in aggregating the information on several cell types.
For example, when several different cancer cell sub-types were identified in the original classification, but the analyst is only interested in how the cancer cells are distributed as a whole.
To that end, we enabled the user to combine cell types, by gradually dragging and dropping the corresponding plots into a drop area on top of the view (\autoref{fig:drag_n_drop}b,c).
The abundances of the dropped cell types are then aggregated as if they were a single cell type and a new distribution is created on-the-fly.

All views in our system are linked and allow cross-selection.
For example, selection one or more lines in a raincloud plot filters the tissue view (\autoref{fig:System}b) to show only the corresponding samples, with the cell type corresponding to the raincloud plot emphasized (\textbf{T4}).
Further, these samples are also highlighted in the other raincloud plots, for example to verify whether a sample that is an outlier for one cell type also shows different behavior for other types (\textbf{T3}).
To ensure that outliers in one cohort are not occluded by samples of the other cohort, the user can select to fade out one of the cohorts (\textbf{T3}).

\subsection{Comparison Based on Cellular Microenvironments}\label{spatial}

The comparison of the cohorts based on their spatial interactions patterns, as indicated in task \textbf{T2}, is performed in two steps.
The first step is to gain a global overview and compare the cohorts based on pairwise co-occurrences of cell types (\textbf{T2.a}).
In the second step, the analyst can go into detail, explore and built specific, detailed microenvironments, consisting of an arbitrary number of cell types, and compare the distribution of these microenvironments among the two cohorts (\textbf{T2.b}).
Throughout this process, we allow locating the identified microenvironments with the actual tissue images (\textbf{T4}) and in the second step, samples that are outliers in their cohort, according to the created microenvironment can be identified (\textbf{T3}).

\subsubsection{Pairwise Overview}\label{overview}

\begin{figure}[t!]
 \centering
 \includegraphics[width=0.9\linewidth]{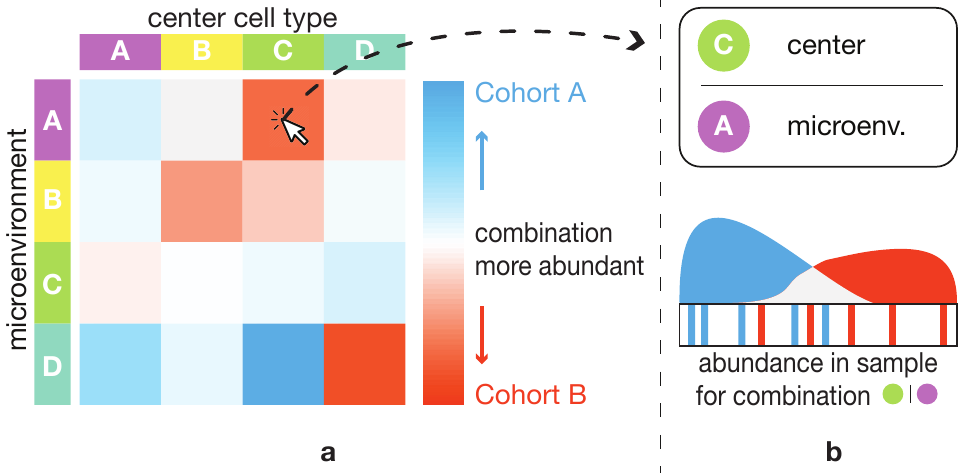}
 \vspace{-3mm}
 \caption{\textbf{Overview of cell type co-localization patterns.} The heatmap (a) explicitly encodes differences in the abundance of pairwise combinations of cell types in the two cohorts. Clicking on one of the combinations sets this combination in the detail view (b), showing the distribution of samples according to the abundance of this combination.
 \vspace{-5mm}
}
 \label{fig:overview_heatmap}
\end{figure}

Following \mimc{}~\cite{Somarakis2019}, we define the microenvironment of a cell, based on a user-defined distance as explained in \autoref{input_data}.
We then compute the frequency for each cell type to occur in each other cell type's microenvironment throughout the cohort.
For a detailed description we refer to our previous work~\cite[Section~4.3]{Somarakis2019}.
The result of this process is a directed and weighted graph, where each node represents a cell type and the link between two nodes defines the frequency of the target node appearing in the microenvironment of the source node.
In \mimc{}, we visualize this frequency graph as a heatmap.
Here, instead of showing the frequencies $F$, we compute the signed differences $D$ in frequency between the two cohorts $C_A$ and $C_B$. 
$D_t(C_A,C_B) = F(C_A) - F(C_B)$.
We encode $D$ using color based on the same heatmap layout,illustrated in \autoref{fig:overview_heatmap}.
The vertical axis shows the cell type of interest and the horizontal axis the cell types in the microenvironments.
A large positive value indicates that the combination exists predominantly in Cohort A, while a large negative value means the combination predominantly exists in Cohort B.
Based on this, we define a simple color map using the same colors previously assigned to the two cohorts and map the maximum absolute value $max(|D_t|)$ to the color assigned to Cohort A (i.e. blue) and $-max(|D_t|)$ to the color assigned to Cohort B (i.e. orange).
Using the same concept of blending between the two colors, described in \autoref{non-spatial}, the middle of this colormap, corresponding to $D_t = 0$, will be a neutral light-grey, indicating both cohorts exhibit similar abundance of the given combination (compare \autoref{fig:overview_heatmap}).

During one of the case studies (\autoref{case_study:1}), it became clear that using the relative frequencies, used in \mimc{}~\cite{Somarakis2019} and the required normalization biased the heatmap towards differences in small cell populations.
To counter this issue, we provide the option to compute the heatmap using the separability metrics, also used for sorting the raincloud plots (\autoref{non-spatial}).
As these metrics only provide information on how different the cohorts are, we compute the mean abundance of the given cell type combination for all samples in a cohort and use the sign of the two cohort's difference in combination with the separability metric.

The resulting heatmap effectively shows cell type combinations that differentiate the two cohorts and for which cohort each combination is predominant.
The analyst can now further explore individual combinations by clicking the corresponding box in the heatmap.
Thereby, the corresponding combination is selected and highlighted in the tissue view (\textbf{T3}) and the microenvironment combination tool (\autoref{details}) is pre-populated with the given combination (\autoref{fig:drag_coloc}a) for further analysis.

\subsubsection{Detail Microenvironments}\label{details}

Starting with the overview of pairwise co-localization patterns, identified with the heatmap visualization, the analyst can now in detail explore complex microenvironment structures, based on any cell type combination and link those to individual samples along their position in the distribution of the corresponding cohort.

\begin{figure}[t!]
 \centering
 \includegraphics[width=0.9\linewidth]{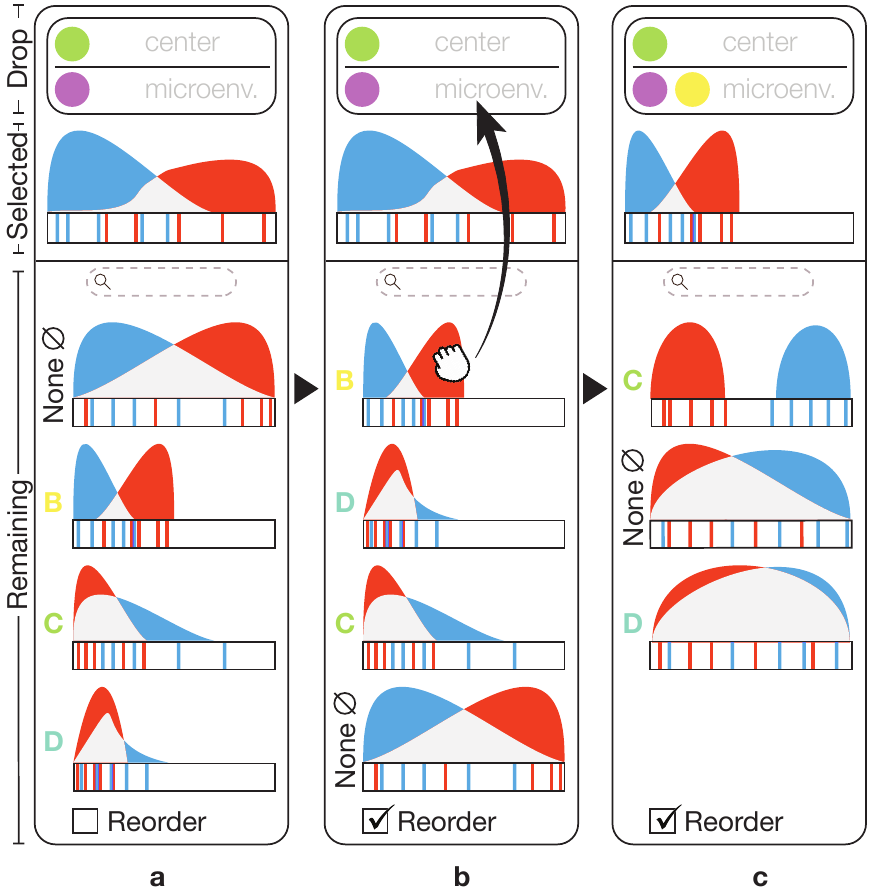}
 \vspace{-3mm}
 \caption{\textbf{Interactive exploration in the detail view.}
 (a) The abundance of the cells fulfilling the cell type pattern in the \emph{Drop} area is illustrated in the \emph{Selected} raincloud plot.
 (b) The raincloud plots are reordered in the \emph{Remaining} area according to their differentiating ability, the user drags the first raincloud plot and drops it in the \emph{Drop} area.
  (c) The dropped raincloud plot replaces the previous one. 
  Also, the \emph{Drop} area and the \emph{Remaining} plots are updated for further exploration. 
 \vspace{-5mm}
}
 \label{fig:drag_coloc}
\end{figure}

In \mimc~\cite{Somarakis2019}, we used a simple glyph to enable the visual exploration of all the existing unique microenvironments in a sample.
Here, the focus is on comparing two cohorts with regard to specific microenvironments, that potentially have already been identified as interesting in a previous analysis of the individual cohorts.
Therefore, instead of showing all the existing unique microenvironments, the user can compare the two cohorts based on a specific pattern of spatial interactions.
To enable the user to interactively define such a pattern, we utilize an interactive visual query system~\cite{Shneiderman1994}, similar to the one presented in Polaris~\cite{Stolte02polaris:a} and further explained by Heer et al.~\cite{heer2012interactive}.
The comparison of the two cohorts then happens with the same raincloud plots introduced in \autoref{non-spatial} but instead of the abundance of a single cell type the plot now displays the abundance of the queried microenvironment.

In practice, the analyst would typically start with a combination of two cell types picked from the heatmap.
This simple microenvironment is illustrated on top of the detail view as illustrated in \autoref{fig:drag_coloc}a, where it is divided into the cell type of interest in the center of the microenvironment (i.e., cell type A, green circle, \autoref{fig:drag_coloc}a) and the microenvironment (i.e., cell type B, purple circle, \autoref{fig:drag_coloc}a).
For the remainder of the paper we will denote microenvironments as \microenv{CBgreen}{CBpurple}, where the circle(s) to the left of the vertical line represents the center cells combined with \textit{or} type and the circle(s) to the right the microenvironment combined with \textit{and}.
I.e., a cell from either of the types left of the line must appear in the center and all the types to the right must appear in the surrounding of this cell.
Below this (\textit{Selected}, \autoref{fig:drag_coloc}a) we show the raincloud plot corresponding to the abundance of all microenvironments with at least the selected combination of cell types.
Finally, further below (\textit{Remaining}, \autoref{fig:drag_coloc}a) we depict the raincloud plots corresponding to the combination of the defined microenvironment plus any of the remaining cell types (here \microenv{CBgreen}{CBpurple}~\microempty,{ } \microenv{CBgreen}{CBpurple}~\envtype{CByellow},{ } \microenv{CBgreen}{CBpurple}~\envtype{CBgreen},{ } \microenv{CBgreen}{CBpurple}~\envtype{CBturqoise}).
The example in \autoref{fig:drag_coloc}a starts with \textit{None} (indicated as \microempty).
At first glance, it might seem surprising that the corresponding raincloud plot is different from the initial plot above it.
\textit{None}, here means that no other additional cell type must exist in the microenvironment, whereas the initial plot shows all microenvironments that at least contain the given types.
We denote this as \microenv{CBgreen}{CBpurple}~\microempty.
Below the \textit{None} plot the remaining combinations are shown with the resulting raincloud plots.
As described in \autoref{non-spatial}, these plots can be ordered according to how strongly the corresponding microenvironment separates the two cohorts.
\autoref{fig:drag_coloc}b illustrates the example after reordering.
With this information the analyst can now continue exploring the microenvironments, for example by dragging the plot corresponding to cell type B (yellow) to the drop area, creating \microenv{CBgreen}{CBpurple}~\envtype{CByellow}, (\autoref{fig:drag_coloc}b).
As the original plot already corresponded to the new microenvironment, we can now simply replace the ``Selected'' plot with the dragged plot (\autoref{fig:drag_coloc}c).
The remaining raincloud plots (\microenv{CBgreen}{CBpurple}~\envtype{CByellow}~\envtype{CBgreen},{ } \microenv{CBgreen}{CBpurple}~\envtype{CByellow}~\microempty,{ } \microenv{CBgreen}{CBpurple}~\envtype{CByellow}~\envtype{CBturqoise}) are re-computed on-the-fly and shown below.
Following this procedure, the user can progressively explore all interesting cell type combinations and evaluate their ability to discriminate the two cohorts and as such their potential as biomarkers.

As described in \autoref{non-spatial}, the raincloud plots make it easy to identify samples that are outliers in their corresponding cohort (\textbf{T3}).
Further, we provide the same linking and brushing features for selecting samples, as described in \autoref{non-spatial}, to link the microenvironment patterns to the tissue view (\textbf{T4}).

\subsection{Tissue View}\label{tissue_view}

\begin{figure}[t!]
 \centering
 \includegraphics[width=0.85\linewidth]{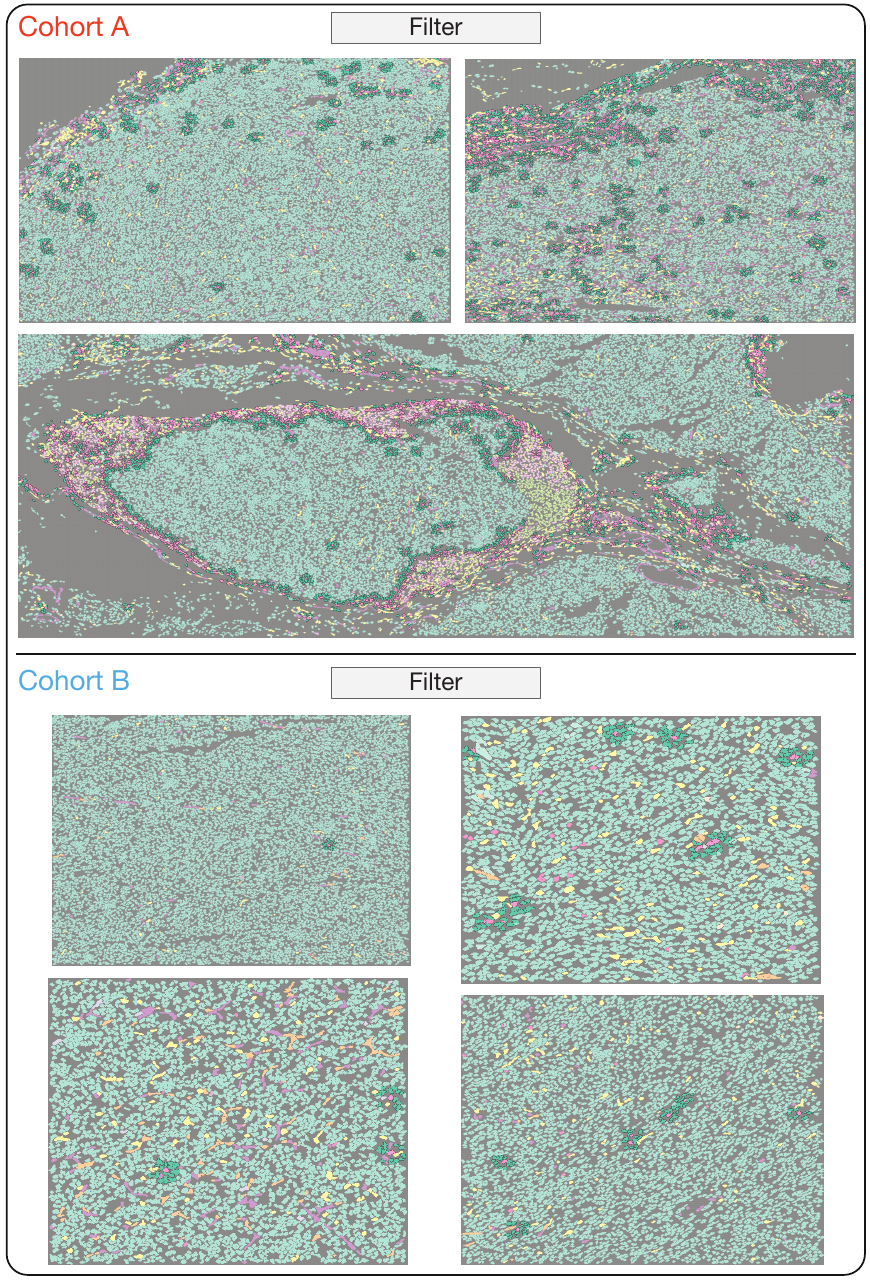}
 \vspace{-2mm}
 \caption{\textbf{Tissue view,} highlighting a spatial interaction fading out the non-selected tissue structures.
 In the tissue samples of Cohort A, the spatial interactions form a compact structure, whereas the spatial interaction of Cohort B tissue samples are distributed all over the samples.%
 \vspace{-3mm}
}
 \label{fig:instances}
\end{figure}

In \autoref{input_data} we have described the importance of enabling the linking of any finding to its spatial location (\textbf{T4}).
Therefore, we provide the tissue view (\autoref{fig:instances}), which shows the original segmented images and, linked to the other views, allows the inspection of selected cell types or microenvironments in the corresponding samples and their spatial context.
The tissue view shows the images using color-coding for the different cell types.
As we only consider the labeled segmentations as input (\autoref{input_data}), we use a categorical colormap to assign a color to each label and thus cell type.
We have chosen the qualitative \textit{12 class Set 3} from \href{https://colorbrewer2.org/#type=qualitative&scheme=Set3&n=12}{colorbrewer}~\cite{brewer1994} and have excluded blue and orange hues to avoid interference with the cohort colors.
Colors are initially assigned based on the order of the cell type labels, but we allow the user to assign them manually by clicking on a cell type label.
As typical studies have more cell types than the available ten colors, they can assign the same color to semantically grouped types.
We then automatically adjust the saturation of hues that were selected multiple times to enable differentiation.
While not described in detail in previous sections, this color scheme is used throughout the application to represent the different cell types and allow for easy mental linking between views.
We have previously used a similar color scheme in \mimc~\cite{Somarakis2019}.
To enable comparison between the cohorts, we divide the tissue view into two parts, one for each cohort.
The name and color corresponding to the cohorts is shown on top of each view (\autoref{fig:instances}).


As described before, all views are linked.
Therefore, the tissue view can be filtered to only show samples selected in other views.
Further, selecting cell types or microenvironments in other views highlights them in the images by fading non-selected structures out, resulting in a light-grey for all unselected areas (\autoref{fig:instances}).
Moreover, the tissue view supports zooming and panning across tissue samples to further assist the exploration of the (highlighted) tissue areas.

\subsection{Implementation}\label{implementation}
As described in \autoref{workflow_users}, our target users are clinical researchers with little programming experience. Therefore, we implemented the described workflow in a stand-alone GUI application.
The application is implemented in MATLAB, as it allowed us to quickly build a stand-alone prototype.
Source code and binaries are available on GitHub~\cite{asom_2020}.

\section{Validation}\label{case_study}
In order to show the effectiveness of our workflow, we conducted three case studies with  collaborators (P1-P3) at Leiden University Medical Center.
P1 was also our main contact during the development of the workflow.
After conducting the case studies and collecting feedback, we invited the collaborators  to participate in the write up of the case-studies, and hence they are all co-authors of this manuscript.
All collaborators acquired their own data with varying biological goals, using two different modalities as indicated in \autoref{table:case_study}.
For the case studies, we 
gave participants a hands-on introduction and answered any questions regarding the tool.
After that, we observed the participants performing their analysis independently and reproduced their workflows for presentation in Sects.~\ref{case_study:1}-\ref{case_study:3}.
As described in \autoref{Workflow}, for all the case studies the segmentation masks and the cell type identification had been performed as a pre-processing step by the participants.
An overview of the study parameters with regard to imaging modality, numbers of samples, and numbers of included cell types is given in \autoref{table:case_study}.
As can be seen, the studies cover three different application areas, contain data from two different modalities, between 20 and 47 samples, and between 12 and 60 cell types.
Finally, we asked the participants, as well as a fourth user of the software (P4, not a co-author of this manuscript), to fill out a short questionnaire (available in the supplemental material) via google forms~\cite{google_forms}.
The questionnaire consists of the ten standard System Usability Scale (SUS) statements~\cite{sus}, an additional nine statements specific to our tool, answered on a 5-point Likert scale, and five questions for open feedback.
The individual plots presented in the case study have been exported directly from our tool and laid out with adjusted labels and annotations for the printout.


\begin{table}[t!]
\centering
\caption{{\textbf{Summary of the case study characteristics.}}}
\renewcommand{\arraystretch}{1.5}
\setlength{\tabcolsep}{5pt}
\newlength\wexp
\settowidth{\wexp}{Samples in Cohort}
\newcolumntype{C}{>{\centering\arraybackslash}p{\dimexpr.5\wexp-\tabcolsep}}
\begin{tabular}{rlcCCc}
\hline
\rowcolor[gray]{.97}&& &  \multicolumn{2}{c}{Samples in Cohort} &  \\ 
\rowcolor[gray]{.97}&\multirow{-2}{*}{Case Study} & \multirow{-2}{*}{Modality} & \textbf{\textcolor{orange}{1}} & \textbf{\textcolor{cyan}{2}} & \multirow{-2}{*}{Cell Types}\\ \hline
P2 & Sarcoma & IMC~\cite{Giesen2014} & 13  & 7 & 12 \\
P1 & Tumor & IMC~\cite{Giesen2014} & 19 & 28 & 60 \\
P3 & Alzheimer's & Vectra~\cite{Nghiem2016} & 12 &  9 & 16 \\ \hline
\end{tabular}%
\label{table:case_study}
\renewcommand{\arraystretch}{1}
 \vspace{-3mm}
\end{table}

\subsection{Case Study I: Synovial Sarcoma (P2)}\label{case_study:1}

Synovial sarcoma is a rare form of cancer.
During the immune response, T-cells infiltrate the sarcomas.
Previous work has shown that synovial sarcomas can have areas with abundant T-cell infiltration (\textit{hot areas}) and areas with very little T-cell infiltration (\textit{cold areas}), in the same tumor\cite{Luk2018}.
The goal of this case study was to explore differences in the immune cell composition between these two types of areas.
A total of 20 areas from 7 different tumors were imaged, of which 7 were cold (\textit{Cold Cohort}, blue) and 13 were hot (\textit{Hot Cohort}, orange).
The size of the samples varied, with the number of cells in each image ranging from $2,678$ to $23,774$ cells.
In the pre-processing step, cells were segmented and 12 different cell types were identified, based on the original data.
While the number of cell types is relatively low, they cover a large range of available types, with rather coarse specificity.

\begin{figure}[b!]
 \centering
 \vspace{-3mm}
 \includegraphics[width=0.9\linewidth]{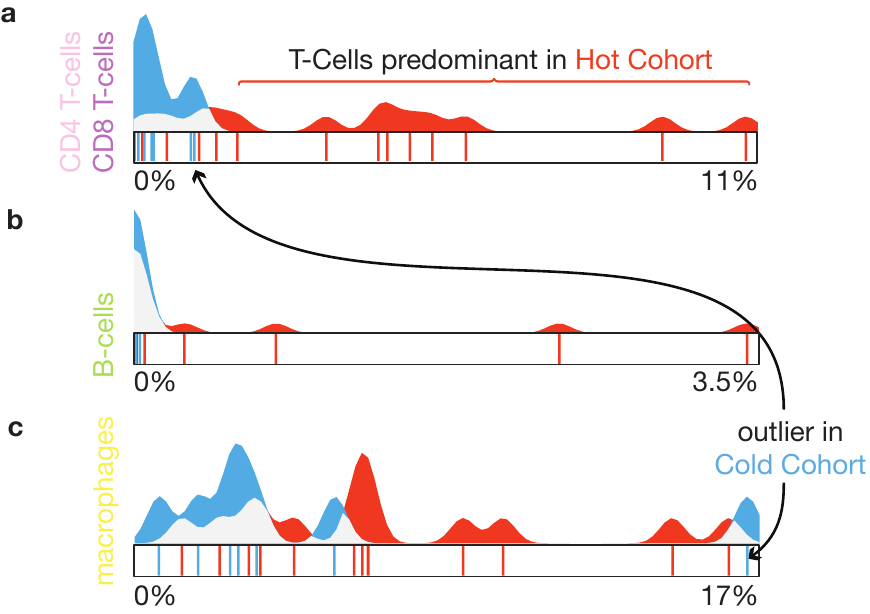}
 \vspace{-3mm}
 \caption{\textbf{Raincloud plots} for combined CD4 and CD8 T-cells (a), B-cells (b), and macrophages (c). An outlier for macrophages in the cold cohort is clearly visible in (c).
 Selecting it showed it also contained slightly more T-cells than other samples in the cold cohort (a).}
 \label{fig:case_study_1_T_M}
\end{figure}
\begin{figure}[b!]
 \centering
 \vspace{-6mm}
 \includegraphics[width=0.9\linewidth]{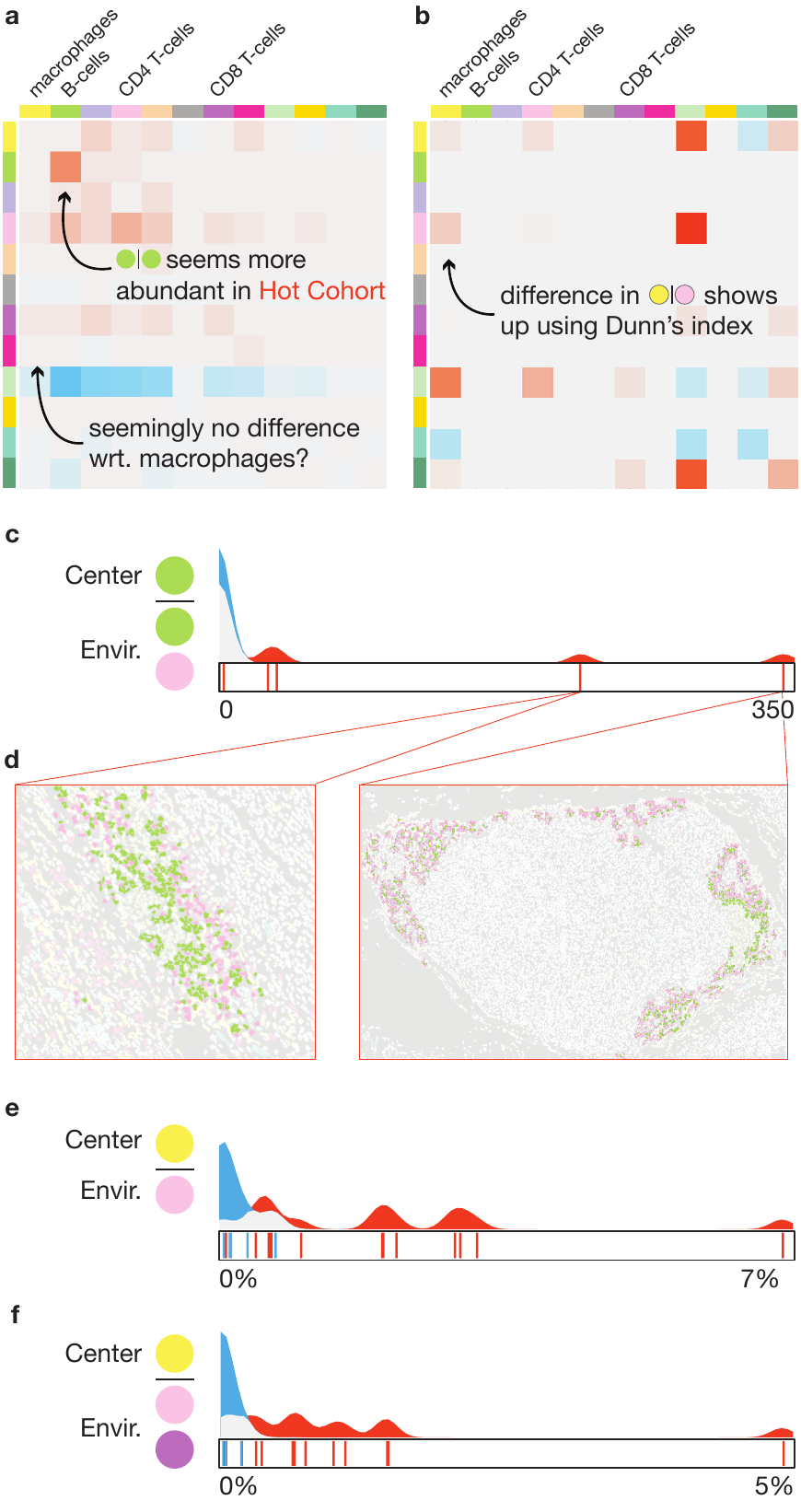}
 \vspace{-2mm}
 \caption{\textbf{Multi-cellular microenvironment cohort comparison.} 
 (a) A heatmap depicting the difference of the amount of  pairwise spatial interaction between two cohorts normalized according to the abundance of each cell type.
 (b) A heatmap depicting the Dunn index for the samples of each cohort for each pairwise co-localization pattern.
 (c) The amount of B-cells having in their microenvironment B-cells and CD4 T-cells, depicting that the occured differentiation in (a) was due to the two outlier samples, which exist in a tertiary lymphoid structure, an interesting biological structure (d).
 The amount of macrophages having in their microenvironment CD4 T-cells (e) and CD8 T-cells (f).
}
 \label{fig:case_study_1_B_cells}
\end{figure}

\subsubsection{Cell Type Abundance}\label{case_study:1_1}

In the first step of the analysis the expert was mostly interested in identifying cell type(s) that differentiate the cohorts, matching \textbf{T1} of our task analysis.
Given the large variation in the number of cells per sample, he used the relative cell type abundance for comparison.
First, he wanted to explore the uniformity of each cohort.
As indicated above, the samples were sorted into the two cohorts based on the infiltration of \textit{T-cells} in the contained tumor tissue.
Consequently the T-cells should exist predominantly in the Hot Cohort.
As a first step, the expert wanted to verify this using the system.
As there are two different types of T-cells in the dataset (\textit{CD4} \envtype{CBpink} and \textit{CD8 T-cells} \envtype{CBpurple}) he first queried for these two cell types and created a combined raincloud plot by dragging the CD4 T-cell and CD8 T-cell plots to the combined drop area (\autoref{non-spatial}).
The resulting combined plot (\autoref{fig:case_study_1_T_M}a) confirmed that T-cells were largely non-existent in all seven samples of the Cold Cohort (blue peak close to 0, \autoref{fig:case_study_1_T_M}a) but more widely distributed in the Hot Cohort (even spread of the orange distribution, \autoref{fig:case_study_1_T_M}a).
After navigating among the plots, he discovered the raincloud plot for \textit{B-cells} \envtype{CBgreen} (\autoref{fig:case_study_1_T_M}b).
This plot caught the expert's interest.
Even though most samples from both cohorts hardly contain any B-cells, there are a few samples in the Hot Cohort that contain some B-cells, indicated by the orange lines to the right of the plot in \autoref{fig:case_study_1_T_M}b.
Given the generally low values, approximately 3 percent, even for the sample with the largest abundance, the expert decided to not further investigate these samples at this point and proceeded with other cell types.
Therefore, he ordered the raincloud plots according to the Dunn's index~\cite{Bezdek1995}.
The first plot illustrating \textit{macrophages} \envtype{CByellow} showed a pattern similar to the T-cells (\autoref{fig:case_study_1_T_M}c).
Strikingly, there is an outlier (\textbf{T3}) clearly visible in the plot (highlight in \autoref{fig:case_study_1_T_M}c).
The corresponding sample from the Cold Cohort consists of over $16\%$ macrophages, compared to no more than $5\%$ for all other samples of the same cohort.
Selecting the corresponding line in the plot also revealed that this sample has the highest abundance of T-cells in this cohort (though only at around $1\%$ of cells in this sample).

At this point, the expert was curious whether the microenvironments of the macrophages and B-cells could provide further clues on differentiating factors between and within the cohorts.

\comment{
Also, the white part was restricted to the left-most part indicating some samples from Cohort 2 (blue) depleted in macrophages.
Also, selecting the lines which represent each sample in the raincloud plot sequentially he observed a correlation among the amount of macrophages and T-cells, with an exception of a sample, which was abundant in macrophages, but depleted in T-cells (dashed line \autoref{fig:case_study_1_T_M}b).
\textcolor{red}{Because hot and cold areas from the same tumors were incorporated, the expert concluded that there may be a spatial interaction within the tumor between T-cells and macrophages.}
}

\comment{Moreover, because of the high abundance of T-cells in Cohort 2 he wanted to examine whether there are also B-cells in these samples, as together they form a tertiary lymphoid structure.
An important structure for the immune system~\cite{}.
However, from the raincloud plot of the B-cells (green) only five samples, belonging in Cohort 2, had B-cells populations.
Moreover, changing the mode to absolute values figured out than only two samples exceeded the defined from the expert threshold of 100 cells.
Selecting the corresponding lines of the remaining samples from the raincloud plot, the tissue samples were highlighted in the tissue view.
From the tissue view he could easily observe that the B-cells (green) form along with T-cells (pink) an indicative structure in both tissue samples illustrated in \autoref{fig:case_study_1_B_cells}c.}

\subsubsection{Micorenvironments}\label{case_study:1_2}

The exploration of the differences between the two cohorts, with regard to the contained microenvironments (\textbf{T2}) starts with the overview provided by the difference heatmap (\autoref{fig:case_study_1_B_cells}a).
The difference heatmap (\autoref{fig:case_study_1_B_cells}a) indicated that combinations of B-cells and B-cells \microenv{CBgreen}{CBgreen} and B-cells and T-cells \microenv{CBgreen}{CBpink} were more prevalent in the Hot Cohort (highlighted orange boxes).
With this information, the expert created the combined mircoenvironment \microenv{CBgreen}{CBgreen}~\envtype{CBpink} using the drag and drop interface.
The corresponding raincloud plot showed two clear outliers in the Hot Cohort showing a larger abundance of this combination (\autoref{fig:case_study_1_B_cells}c).
Using the linked tissue view, the expert could highlight the microenvironments in the corresponding samples (\autoref{fig:case_study_1_B_cells}d).
The expert observed that the highlighted microenvironments were mostly present in so-called tertiary lymphoid structures~\cite{Luk2018}.
While not directly relevant for the cohort comparison, he noted the two outlier samples for later detailed inspection in his standard workflow.

In the previous step, the expert had also identified macrophages \envtype{CByellow} for further exploration.
Curiously, the heatmap did not show any strong differences between the two cohorts with regard to the microenvironments of this cell type.
After the case study, we analyzed the data and came to the conclusion that the normalization applied to create the heatmap (\autoref{overview}) strongly biased the heatmap in favor of small cell populations such as the B-cells in this study (\autoref{case_study:1_1}).
As a result, we added the option to use the same cluster separation metrics used for sorting the raincloud plots according to their power to separate the cohorts for the heatmap as described in \autoref{overview}.
\autoref{fig:case_study_1_B_cells}b shows the heatmap using the Dunn's index as an example.
Here, the \microenv{CByellow}{CBpink} microenvironment is more clearly visible, while the small values of the B-Cell microenvironments are suppressed.
The expert selected the corresponding box from the heatmap and examined the distribution of the samples for each cohort in the detail view. 
The blue area around zero (\autoref{fig:case_study_1_B_cells}e) indicated the absence of \microenv{CByellow}{CBpink} microenvironment in the Cold Cohort, verifying the heatmap findings.
Then, the expert having already identified the correlation among CD8 T-cells and macrophages navigated among the plots of the ``Remaining'' area of the detail view and located the CD8 T-cell raincloud plot.
The addition of CD8 T-cells in the microenvironment of macrophages \microenv{CByellow}{CBpurple}~\envtype{CBpink} further differentiated the two cohorts, shown by the restriction of the blue area to almost zero (\autoref{fig:case_study_1_B_cells}f).
Even the strong outlier in the Cold Cohort that contained the largest amount macrophages of all samples did not show any significant co-localization of macrophages and T-cells.
On the other hand, several samples in the Hot Cohort showed significant amounts of both combinations.
Therefore, the expert concluded that both T-cell sub-types seems to better differentiate the hot and cold tumor areas, than their one-to-one spatial interaction or even their abundances.
\begin{figure}[b!]
 \centering
 \includegraphics[width=0.9\linewidth]{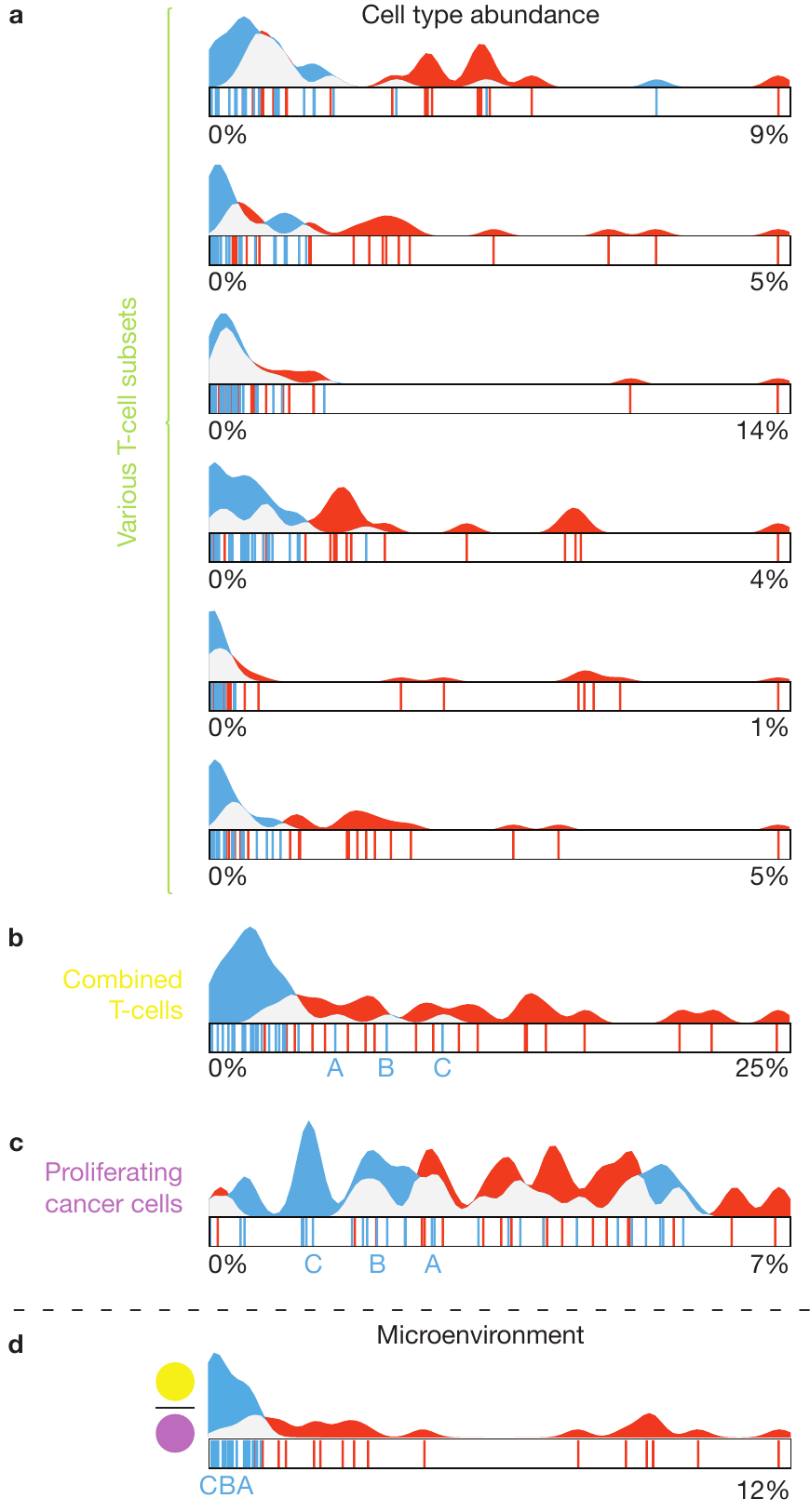}
 \caption{
\textbf{Raincloud plots} for various T-cell subsets (a), the aggregated plot combined from those subsets (b), and proliferating cancer cells (c). (d) shows the amount of the aggregated T-cells with proliferating cancer cells in their microenvironment.
Even though the samples A-C, of the Metastatic Cohort, had a significant amount of T-cells and proliferating cancer cells (b,c) they did not spatially interact (d).}
 \label{fig:case_study_2_complete}
\end{figure}\label{case_study:2}

\subsection{Case Study II: Tumor Metastasis (P1)}

In this case study, the expert wanted to explore the differences in the cellular microenvironments of tumors with different clinical characteristics.
In particular, she had acquired a data set, consisting of a total of $47$ images taken from different tumor samples.
Based on other clinical parameters she divided the set in two cohorts.
The first one contains $19$ images of non-metastatic tumors (\textit{Non-Metastatic Cohort}, orange), the second $28$ images of metastatic tumors (\textit{Metastatic Cohort}, blue).
She had segmented the images in a pre-processing step and identified $60$ different cell types, among a total of $393,727$ cells.

\subsubsection{Cell Type Abundance}\label{case_study:2_1}

First, the expert was interested to discover cell type(s) which exist predominantly in one of the cohorts.
Given the large amount of cell types, she ordered the raincloud plots according to the Silhouette metric in descending order, to assist her exploration.
The first few plots consisted mostly of different subsets of T-cells, which had been defined in great detail in the preprocessing step.
All of the corresponding plots showed a similar pattern of very small abundances for the Metastatic Cohort, indicated by a large blue peak to the left of the plot and a varying, but generally larger abundance in the Non-Metastatic Cohort.
Searching for all cell types containing ``T-cell'' in their label showed a similar pattern for all of the remaining types (\autoref{fig:case_study_2_complete}a).
This pattern is not completely surprising, as T-cells are a major factor in the immune response to cancer.
For further exploration, in particular the relation of the identified T-cells to cancer cells, the expert aggregated all T-cell subsets using the drag and drop interface.
The resulting raincloud plot (\autoref{fig:case_study_2_complete}b) confirmed that the T-cells \envtype{CByellow} clearly differentiate the two cohorts.
There were, however, three samples from the Metastatic Cohort visible (blue lines, labeled A,B,C in \autoref{fig:case_study_2_complete}b) that showed a somewhat increased abundance compared to the remaining samples in that cohort.
Next, the expert was interested, whether the increased amount of T-cells in the Non-Metastatic Cohort would correlate to differences in contained tumor cells.
The expert searched for ``tumor'', to bring up the raincloud plot, corresponding to \textit{Proliferating Tumor Cells} \envtype{CBpurple}.
However, as shown in \autoref{fig:case_study_2_complete}c, no clear separation between the two cohorts can be made, based on these cells.
Finally, selecting the three outliers samples (A,B,C) in the T-cell plot did not show a specific differentiation with regard to the tumor cells.

\subsubsection{Micorenvironments}\label{case_study:2_2}

The last findings of \autoref{case_study:2_1} intrigued the interest of the expert to further explore whether the tumor cells are present in the same amounts also in the microenvironment of T-cells.
She quickly combined T-cells and proliferating tumor cells to a microenvironment \microenv{CByellow}{CBpurple} to bring up the corresponding raincloud plot (\autoref{fig:case_study_2_complete}d) in the detail view.
The plot shows a clear differentiation among the two cohorts.
In fact, this combination differentiates the two cohorts even stronger than only the T-cells.
Even for the samples (Samples A,B,C) that showed increased abundance in T-cells, compared to the rest of the Metastatic Cohort, there was only a very small abundance of the \microenv{CByellow}{CBpurple} microenvironment.
This strongly indicates that tumor cells exist in the microenvironment of T-cells in the Non-Metastatic Cohort, whereas in the Metastatic Cohort there is no spatial interaction between tumor and T-cells regardless their abundance.
This lead the expert to hypothesize that the co-localization between the tumor and T-cells needs to be taken into account in tumor analysis, rather than the abundance of T-cells alone.

\subsection{Case Study III: Alzheimer's Disease (P3)}
\label{study_alzheimers}

The accumulation of \textit{amyloid plaques} in the brain is an important characteristic of Alzheimer disease.
These amyloid plaques are infiltrated by microglial cells, the resident immune cells of the brain.
In this final case study, the expert wanted to verify the hypothesis that the microglia cells close to and potentially attacking amyloid plaques are different from the microglia cells in healthy individuals.

The data used in this case study are somewhat different from the first two cases.
The number of samples is comparable.
Here, each sample represents one subject, for a total of 12 patients in the Alzheimer's Cohort (orange) and 9 healthy subjects in the Control Cohort (blue).
However, each subject is described by up to $150$ images, acquired with the Vectra 3.0~\cite{Nghiem2016} machinery.
$16$ different cell types were identified and segmented in the pre-processing step.
The identified cell types consist mostly of different subsets of microglia cells and as a result, the segmentation of the images is rather sparse, containing only in the order of $25$ cells per image, plus the separately segmented amyloid plaques.
As such, the individual images were not as important in this study as in the previous two and the data set only contained aggregated information of cell type abundance and microenvironments for all images per subject.

 \begin{figure}[t!]
 \centering
 \includegraphics[width=0.9\linewidth]{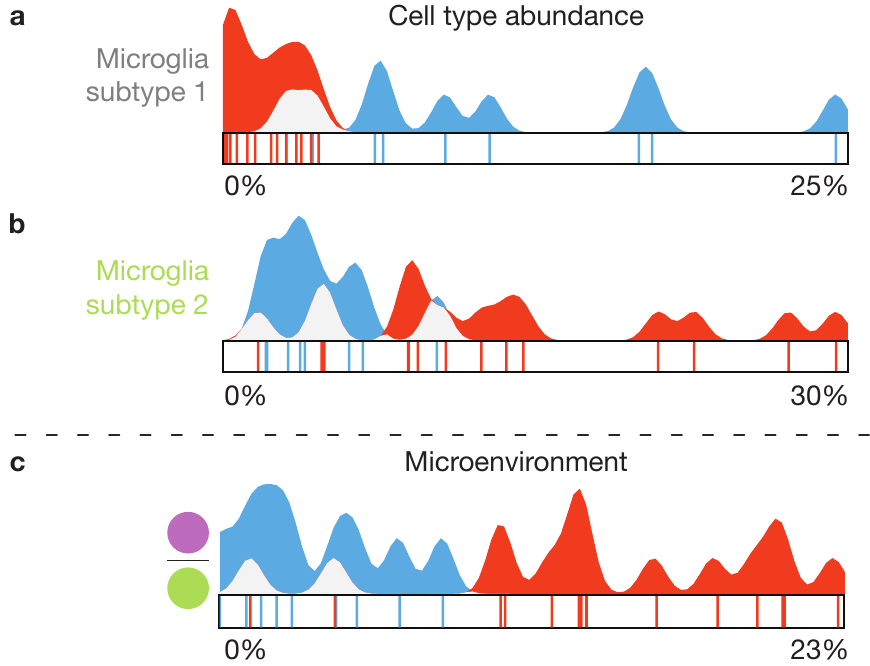}
 \caption{\textbf{Raincloud plots} for microglia subtypes  (a,b), and amoyloid plaques  with microglia  subtype 2 in their microenvironment (c).
 \vspace{-3mm}
}
 \label{fig:case_study_3_complete}
\end{figure}\label{case_study:3}

\subsubsection{Data Analysis}
 
As the experts goal was to verify a specific hypothesis, the data analysis in this study was much more targeted, compared to the rather explorative nature of the previous case studies.
First, he brought up the raincloud plots corresponding to two microglia subtypes with contradictory patterns (\autoref{fig:case_study_3_complete}a,b).
As can be seen in the plots \textit{Subtype 1} \envtype{CBgrey} was prevalent in the Control Cohort (blue), whereas Subtype 2 \envtype{CBgreen} was mostly found in samples of the Alzheimer's Cohort (orange) but there was still some overlap between the samples from the two cohorts.
This differentiation was already an indicator to verify the original hypothesis of the expert.
Going back to the original data, the expert noted that the microglia Subtype 2 did not express two proteins that were expressed by Subtype 1 and hypothesized that these proteins might be suppressed when in the vicinity of the amyloid plaques \envtype{CBpurple} in Alzheimer's disease patients.
Consequently, he brought up the raincloud plot of the corresponding microenvironment \microenv{CBpurple}{CBgreen} (\autoref{fig:case_study_3_complete}c).
Here, the distinction between the two cohorts is even clearer, with only two samples from the Alzheimer's Cohort in the range of the Control Cohort.
The distribution further indicates that Subtype 2 seems to co-localize with amyloid plaques, supporting the generated hypothesis.

\subsection{Feedback}\label{feedback}

After the case studies, we collected feedback from the participants using a short questionnaire (available in the supplemental material) via google forms~\cite{google_forms}.
The questionnaire consists of the ten standard System Usability Scale (SUS) statements~\cite{sus} (Q1--Q10), an additional nine statements specific to our tool (Q11--Q19), answered on a 5-point Likert scale, and five questions for open feedback.
After the case studies, a fourth collaborator started working with the tool. After she got acquainted with it, we asked her to fill out the same questionnaire.

The average SUS-score, based on all four questionnaires was $76.25$ with a standard deviation of $3.23$ resulting in a \textit{good} rating~\cite{susrating}.
In the following we briefly summarize the feedback of the custom block of the questionnaire (Q11--Q19), for the complete set of responses we refer to the supplemental material.
An overview of the responses is provided in \autoref{tab:questionnaire}.
The custom part of the questionnaire is divided into three blocks.
The first block (Q11--Q14) corresponds to the identified tasks (\autoref{id_tasks}).
The second block (Q15--Q18) targets the
interaction with the 
raincloud-based views in the cell abundance and microenvironment exploration.
Finally, in the third block, we ask about general feedback.

With statements Q11--Q14 we queried whether \textbf{T1}--\textbf{T4} (\autoref{id_tasks}) could be carried out efficiently.
(Q11; \textit{The tool allows me to efficiently compare two cohorts, according to the abundance of contained cell types per sample} relates to \textbf{T1}, Q12 to \textbf{T2}, and so on).
Generally, responses were clearly positive with strongly agree (++) or agree (+) with the exception of a neutral (\neutral) response to Q11 and Q12, each.
From the open feedback (Q20: \textit{What functionality was missing to fully accomplish all goals?}) we could gather that participants would like to be able to \qq{correct[ion] cell abundance} with regard to the amount of cells from user-defined area.
Further, \qq{statistical testing of differences found between cohorts} was requested, related to \textbf{T1} and \textbf{T2}.

In Q15--Q18 we were interested whether the raincloud plots were helpful to compare the distributions (Q15, \textbf{T1-T2}) and to find outliers (Q16, \textbf{T3}) as well as whether the drag and drop interaction made it easy to combine cell types (Q17) and build microenvironments (Q18).
Q15--Q17 were overwhelmingly positive, with Q18 getting neutral responses by majority.
The different response to Q17 and Q18 is rather unclear to us, as the interaction for combining cell types and building the detailed microenvironments is essentially the same.
Unfortunately, there is also no further feedback on this in the open part of the questionnaire.

\begin{table}[!t]
\caption{\textbf{Summary of participants’ answers} to statements of our questionnaire on a 5-point Likert scale from \textit{very positive} (++) to \textit{negative} (-). No \textit{very negative} (--) responses were given.}
\renewcommand{\arraystretch}{1.25}
\setlength\tabcolsep{5pt}
\begin{tabular}{r | >{\columncolor[gray]{0.97}}p{0.55cm} p{0.55cm} >{\columncolor[gray]{0.97}}p{0.55cm} p{0.55cm} | >{\columncolor[gray]{0.97}}p{0.55cm} p{0.55cm} >{\columncolor[gray]{0.97}}p{0.55cm} p{0.55cm} | >{\columncolor[gray]{0.97}}p{0.55cm}}
Q  & 11  & 12   & 13    & 14    & 15    & 16    & 17    & 18    & 19    \\ \hline
++ & \qz &      & \qz   & \qp   & \qz   & \qp   & \qp   &       &       \\
+  & \qp & \qd  & \qz   & \qd   & \qz   & \qz   & \qd   & \qp   & \qz   \\
\neutral  & \qp & \qp  &       &       &       & \qp   &       & \qd   &       \\
-  &     &      &       &       &       &       &       &       & \qz   \\ 
\end{tabular}
\label{tab:questionnaire}
\renewcommand{\arraystretch}{1}
\end{table}

In the open feedback we can see that Participant 3 was missing \qq{Within subject distribution of cell types/clusters.}
As described in \autoref{case_study:3}, we had aggregated the very large amount of images in this study to a single dataset per subject.
It might be interesting to provide a hierarchical approach in the future, that allows drilling into these subjects.
Participant 4 mentioned \qq{the option to compare 3 cohorts} as a missing feature in the open feedback.
While we focus on  the comparison between two cohorts this is a possible future extension.

Finally, in the open feedback the \qq{possibility to detect outliers (and directly identify the subject} (\textbf{T3}) was specifically mentioned as a positive aspect.
The link between the abstract views and the actual images (\textbf{T4}) was highlighted by one participant:
\qq{The rainbowplots are really cool, especially because you can go up and down to the images again.}
Particularly positive was a comment by Participant 1, that
\qq{with the tool I already discovered a very nice thing in my existing data!}.

\section{Discussion and Conclusion}\label{conclusion}
We presented a workflow for the interactive visual comparison of two cohorts comprising single-cell \omi{}-data, based on the cell abundance and their cell microenvironments.

The presented case studies contained up to 47 samples and up to nearly 400.000 cells.
Our sorting and filtering options allow effective exploration of datasets of such sizes, however, increasing numbers to hundreds of samples will pose new challenges.
In the Alzheimer's disease case study we accommodated a much larger original dataset (3286 images) by aggregating the information per patient and imaged region to a single larger image, resulting in the  dataset described in \autoref{case_study:3}.
Extending this to a hierarchical approach, facilitating the exploration of such aggregated regions and then individual images within a region might be a worthwhile extension.

At this point, our workflow is focused on two-dimensional images, as our partners currently only acquire such data.
However, image stacks or volumetric measurements are becoming more readily available.
Assuming a three-dimensional definition of microenvironments, the views based on abstract information, such as the raincloud plots and heatmap, would readily adapt to such data.
Extensions to the spatial view, for example by volume rendering, would be necessary to inspect findings in the tissue context.

We have implemented the drag and drop interface to create simple center-neighborhood microenvironments.
Nevertheless, the approach would support more advanced microenvironments through more drop targets, intuitively. For example, the neighborhood could be divided into multiple segments to allow a microenvrionment definition that has cell type A to the left and cell type B to the right of the center cell.
A more traditional user interface, such as checkboxes, to assign cell types to each of those segments would be less flexible and quickly require a large amount of additional user interface elements.

Our workflow is designed to compare two clearly defined separate cohorts such as control vs. disease.
Extending it to support more cohorts, or including more continuous features such as age or trial dose are open questions that certainly warrant future research.
\acknowledgments{%
This work received funding through Leiden University Data Science Research Programme.
B.P.F.Lelieveldt received partial funding from H2020-Marie Skodowska-Curie Action Research and Innovation Staff Exchange (RISE) Grant 644373-PRISAR.
N.F.C.C.de Miranda has received funding from the European Research Council (ERC) under the European Union’s Horizon 2020 research and innovation program (grant agreement No. 852832)
}
\bibliographystyle{abbrv-doi-hyperref}

\bibliography{bibliography}

\begin{thebibliography}{10}

\bibitem{Ali2020}
\href{https://doi.org/10.1038/s43018-020-0026-6}{H.~R. Ali, H.~W. Jackson,
  V.~R.~T. Zanotelli, E.~Danenberg, J.~R. Fischer, H.~Bardwell, E.~Provenzano,
  O.~M. Rueda, S.-F. Chin, S.~Aparicio, C.~Caldas, and B.~Bodenmiller}.
\newblock \href{https://doi.org/10.1038/s43018-020-0026-6}{Imaging mass
  cytometry and multiplatform genomics define the phenogenomic landscape of
  breast cancer}.
\newblock \href{https://doi.org/10.1038/s43018-020-0026-6}{{\em Nature
  Cancer}}, \href{https://doi.org/10.1038/s43018-020-0026-6}{1(2):163--175},
  \href{https://doi.org/10.1038/s43018-020-0026-6}{2020}.
  \href{https://doi.org/10.1038/s43018-020-0026-6}
{doi: {{%
10\hspace{.1pt}\discretionary{.}{%
}{.}\hspace{.4pt}1038\discretionary{/}{%
}{/}s43018\discretionary{%
}{-}{-}020\discretionary{%
}{-}{-}0026\discretionary{%
}{-}{-}6}}}


\bibitem{Allen2018}
\href{https://doi.org/10.12688/wellcomeopenres.15191.1}{M.~Allen, D.~Poggiali,
  K.~Whitaker, T.~R. Marshall, and R.~Kievit}.
\newblock \href{https://doi.org/10.12688/wellcomeopenres.15191.1}{Raincloud
  plots: a multi-platform tool for robust data visualization [version 1; peer
  review: 2 approved]}.
\newblock \href{https://doi.org/10.12688/wellcomeopenres.15191.1}{{\em Wellcome
  open research}}, \href{https://doi.org/10.12688/wellcomeopenres.15191.1}{4},
  \href{https://doi.org/10.12688/wellcomeopenres.15191.1}{2019}.
  \href{https://doi.org/10.12688/wellcomeopenres.15191.1}
{doi: {{%
10\hspace{.1pt}\discretionary{.}{%
}{.}\hspace{.4pt}12688\discretionary{/}{%
}{/}wellcomeopenres\hspace{.1pt}\discretionary{.}{%
}{.}\hspace{.4pt}15191\hspace{.1pt}\discretionary{.}{%
}{.}\hspace{.4pt}1}}}


\bibitem{asom_2020}
\href{https://doi.org/10.5281/zenodo.3885814}{asom}.
\newblock \href{https://doi.org/10.5281/zenodo.3885814}{biovault/spaceco
  v.1.0.0}.
\newblock \href{https://doi.org/10.5281/zenodo.3885814}{{\em Zenodo}},
  \href{https://doi.org/10.5281/zenodo.3885814}{2020}.
  \href{https://doi.org/10.5281/zenodo.3885814}
{doi: {{%
10\hspace{.1pt}\discretionary{.}{%
}{.}\hspace{.4pt}5281\discretionary{/}{%
}{/}zenodo\hspace{.1pt}\discretionary{.}{%
}{.}\hspace{.4pt}3885814}}}


\bibitem{susrating}
A.~Bangor, P.~Kortum, and J.~Miller.
\newblock Determining what individual {SUS} scores mean: Adding an adjective
  rating scale.
\newblock {\em Journal of Usability Studies}, 3:114--1234, 1996.

\bibitem{Basole2015}
\href{https://doi.org/10.1145/2836034.2836040}{R.~C. Basole, H.~Park, M.~Gupta,
  M.~L. Braunstein, D.~H. Chau, M.~Thompson, V.~Kumar, R.~Pienta, and
  M.~Kahng}.
\newblock \href{https://doi.org/10.1145/2836034.2836040}{A visual analytics
  approach to understanding care process variation and conformance}.
\newblock \href{https://doi.org/10.1145/2836034.2836040}{In {\em Proceedings of
  the Workshop on Visual Analytics in Healthcare (VAHC)}},
  \href{https://doi.org/10.1145/2836034.2836040}{2015}.
  \href{https://doi.org/10.1145/2836034.2836040}
{doi: {{%
10\hspace{.1pt}\discretionary{.}{%
}{.}\hspace{.4pt}1145\discretionary{/}{%
}{/}2836034\hspace{.1pt}\discretionary{.}{%
}{.}\hspace{.4pt}2836040}}}


\bibitem{Bezdek1995}
\href{https://doi.org/10.1109/ANNES.1995.499469}{J.~C. Bezdek and N.~R. Pal}.
\newblock \href{https://doi.org/10.1109/ANNES.1995.499469}{Cluster validation
  with generalized dunn's indices}.
\newblock \href{https://doi.org/10.1109/ANNES.1995.499469}{In {\em Proceedings
  of Artificial Neural Networks and Expert Systems (ANNES)}},
  \href{https://doi.org/10.1109/ANNES.1995.499469}{pp. 190--193},
  \href{https://doi.org/10.1109/ANNES.1995.499469}{1995}.
  \href{https://doi.org/10.1109/ANNES.1995.499469}
{doi: {{%
10\hspace{.1pt}\discretionary{.}{%
}{.}\hspace{.4pt}1109\discretionary{/}{%
}{/}ANNES\hspace{.1pt}\discretionary{.}{%
}{.}\hspace{.4pt}1995\hspace{.1pt}\discretionary{.}{%
}{.}\hspace{.4pt}499469}}}


\bibitem{v-plots2020}
\href{https://doi.org/10.1111/cgf.14002}{M.~Blumenschein, L.~J. Debbeler, N.~C.
  Lages, B.~Renner, D.~A. Keim, and M.~El-Assady}.
\newblock \href{https://doi.org/10.1111/cgf.14002}{v-plots: Designing hybrid
  charts for the comparative analysis of data distributions}.
\newblock \href{https://doi.org/10.1111/cgf.14002}{{\em Computer Graphics
  Forum}}, \href{https://doi.org/10.1111/cgf.14002}{39(3):565--577},
  \href{https://doi.org/10.1111/cgf.14002}{2020}.
  \href{https://doi.org/10.1111/cgf.14002}
{doi: {{%
10\hspace{.1pt}\discretionary{.}{%
}{.}\hspace{.4pt}1111\discretionary{/}{%
}{/}cgf\hspace{.1pt}\discretionary{.}{%
}{.}\hspace{.4pt}14002}}}


\bibitem{Brehmer2013}
\href{https://doi.org/10.1109/TVCG.2013.124}{M.~Brehmer and T.~Munzner}.
\newblock \href{https://doi.org/10.1109/TVCG.2013.124}{A multi-level typology
  of abstract visualization tasks}.
\newblock \href{https://doi.org/10.1109/TVCG.2013.124}{{\em IEEE Transactions
  on Visualization and Computer Graphics}},
  \href{https://doi.org/10.1109/TVCG.2013.124}{19(12):2376--2385},
  \href{https://doi.org/10.1109/TVCG.2013.124}{2013}.
  \href{https://doi.org/10.1109/TVCG.2013.124}
{doi: {{%
10\hspace{.1pt}\discretionary{.}{%
}{.}\hspace{.4pt}1109\discretionary{/}{%
}{/}TVCG\hspace{.1pt}\discretionary{.}{%
}{.}\hspace{.4pt}2013\hspace{.1pt}\discretionary{.}{%
}{.}\hspace{.4pt}124}}}


\bibitem{brewer1994}
\href{https://doi.org/10.1559/152304003100010929}{C.~A. Brewer, G.~W. Hatchard,
  and M.~A. Harrower}.
\newblock \href{https://doi.org/10.1559/152304003100010929}{{ColorBrewer} in
  print: A catalog of color schemes for maps}.
\newblock \href{https://doi.org/10.1559/152304003100010929}{{\em Cartography
  and Geographic Information Science}},
  \href{https://doi.org/10.1559/152304003100010929}{30(1):5--32},
  \href{https://doi.org/10.1559/152304003100010929}{2003}.
  \href{https://doi.org/10.1559/152304003100010929}
{doi: {{%
10\hspace{.1pt}\discretionary{.}{%
}{.}\hspace{.4pt}1559\discretionary{/}{%
}{/}152304003100010929}}}


\bibitem{sus}
J.~Brooke.
\newblock {SUS}: a ``quick and dirty'' usability scale.
\newblock In P.~W. Jordan, B.~Thomas, B.~A. Weerdmeester, and I.~L. McClelland,
  eds., {\em Usability Evaluation in Industry}, pp. 189--194. Taylor and
  Francis, 1996.

\bibitem{Cibulski2016}
L.~Cibulski and B.~Preim.
\newblock Visual analytics support for analysis of cohort study data:
  Requirements and concepts.
\newblock Technical report, Otto-Von-Guericke University Magdeburg, 2016.

\bibitem{conesa2019making}
\href{https://doi.org/10.1038/s41597-019-0258-4}{A.~Conesa and S.~Beck}.
\newblock \href{https://doi.org/10.1038/s41597-019-0258-4}{Making multi-omics
  data accessible to researchers}.
\newblock \href{https://doi.org/10.1038/s41597-019-0258-4}{{\em Scientific
  data}}, \href{https://doi.org/10.1038/s41597-019-0258-4}{6(1):1--4},
  \href{https://doi.org/10.1038/s41597-019-0258-4}{2019}.
  \href{https://doi.org/10.1038/s41597-019-0258-4}
{doi: {{%
10\hspace{.1pt}\discretionary{.}{%
}{.}\hspace{.4pt}1038\discretionary{/}{%
}{/}s41597\discretionary{%
}{-}{-}019\discretionary{%
}{-}{-}0258\discretionary{%
}{-}{-}4}}}


\bibitem{correll2018}
\href{https://doi.org/10.1109/TVCG.2018.2864907}{M.~Correll, M.~Li,
  G.~Kindlmann, and C.~Scheidegger}.
\newblock \href{https://doi.org/10.1109/TVCG.2018.2864907}{Looks good to me:
  Visualizations as sanity checks}.
\newblock \href{https://doi.org/10.1109/TVCG.2018.2864907}{{\em IEEE
  Transactions on Visualization and Computer Graphics}},
  \href{https://doi.org/10.1109/TVCG.2018.2864907}{25(1):830--839},
  \href{https://doi.org/10.1109/TVCG.2018.2864907}{2019}.
  \href{https://doi.org/10.1109/TVCG.2018.2864907}
{doi: {{%
10\hspace{.1pt}\discretionary{.}{%
}{.}\hspace{.4pt}1109\discretionary{/}{%
}{/}TVCG\hspace{.1pt}\discretionary{.}{%
}{.}\hspace{.4pt}2018\hspace{.1pt}\discretionary{.}{%
}{.}\hspace{.4pt}2864907}}}


\bibitem{Crosetto2015}
\href{https://doi.org/10.1038/nrg3832}{N.~Crosetto, M.~Bienko, and A.~{Van
  Oudenaarden}}.
\newblock \href{https://doi.org/10.1038/nrg3832}{Spatially resolved
  transcriptomics and beyond}.
\newblock \href{https://doi.org/10.1038/nrg3832}{{\em Nature Reviews
  Genetics}}, \href{https://doi.org/10.1038/nrg3832}{16(1):57--66},
  \href{https://doi.org/10.1038/nrg3832}{2015}.
  \href{https://doi.org/10.1038/nrg3832}
{doi: {{%
10\hspace{.1pt}\discretionary{.}{%
}{.}\hspace{.4pt}1038\discretionary{/}{%
}{/}nrg3832}}}


\bibitem{dinkla2017screenit}
\href{https://doi.org/10.1109/TVCG.2016.2598587}{K.~Dinkla, H.~Strobelt,
  B.~Genest, S.~Reiling, M.~Borowsky, and H.~Pfister}.
\newblock \href{https://doi.org/10.1109/TVCG.2016.2598587}{Screenit: Visual
  analysis of cellular screens}.
\newblock \href{https://doi.org/10.1109/TVCG.2016.2598587}{{\em IEEE
  Transactions on Visualization and Computer Graphics}},
  \href{https://doi.org/10.1109/TVCG.2016.2598587}{PP(99):1--1},
  \href{https://doi.org/10.1109/TVCG.2016.2598587}{2017}.
  \href{https://doi.org/10.1109/TVCG.2016.2598587}
{doi: {{%
10\hspace{.1pt}\discretionary{.}{%
}{.}\hspace{.4pt}1109\discretionary{/}{%
}{/}TVCG\hspace{.1pt}\discretionary{.}{%
}{.}\hspace{.4pt}2016\hspace{.1pt}\discretionary{.}{%
}{.}\hspace{.4pt}2598587}}}


\bibitem{Dzyubachyk2013}
\href{https://doi.org/10.1007/s11548-013-0820-z}{O.~Dzyubachyk, J.~Blaas, C.~P.
  Botha, M.~Staring, M.~Reijnierse, J.~L. Bloem, R.~J. {Van Der Geest}, and
  B.~P. Lelieveldt}.
\newblock \href{https://doi.org/10.1007/s11548-013-0820-z}{Comparative
  exploration of whole-body {MR} through locally rigid transforms}.
\newblock \href{https://doi.org/10.1007/s11548-013-0820-z}{{\em International
  Journal of Computer Assisted Radiology and Surgery}},
  \href{https://doi.org/10.1007/s11548-013-0820-z}{8(4):635--647},
  \href{https://doi.org/10.1007/s11548-013-0820-z}{2013}.
  \href{https://doi.org/10.1007/s11548-013-0820-z}
{doi: {{%
10\hspace{.1pt}\discretionary{.}{%
}{.}\hspace{.4pt}1007\discretionary{/}{%
}{/}s11548\discretionary{%
}{-}{-}013\discretionary{%
}{-}{-}0820\discretionary{%
}{-}{-}z}}}


\bibitem{Femino1998}
\href{https://doi.org/10.1126/science.280.5363.585}{A.~M. Femino, F.~S. Fay,
  K.~Fogarty, and R.~H. Singer}.
\newblock \href{https://doi.org/10.1126/science.280.5363.585}{Visualization of
  single rna transcripts in situ}.
\newblock \href{https://doi.org/10.1126/science.280.5363.585}{{\em Science}},
  \href{https://doi.org/10.1126/science.280.5363.585}{280(5363):585--590},
  \href{https://doi.org/10.1126/science.280.5363.585}{1998}.
  \href{https://doi.org/10.1126/science.280.5363.585}
{doi: {{%
10\hspace{.1pt}\discretionary{.}{%
}{.}\hspace{.4pt}1126\discretionary{/}{%
}{/}science\hspace{.1pt}\discretionary{.}{%
}{.}\hspace{.4pt}280\hspace{.1pt}\discretionary{.}{%
}{.}\hspace{.4pt}5363\hspace{.1pt}\discretionary{.}{%
}{.}\hspace{.4pt}585}}}


\bibitem{google_forms}
Google forms.
\newblock \url{https://www.google.com/forms/about/}.
\newblock Accessed: 2020-04-20.

\bibitem{Giesen2014}
\href{https://doi.org/10.1038/nmeth.2869}{C.~Giesen, H.~A. Wang, D.~Schapiro,
  N.~Zivanovic, A.~Jacobs, B.~Hattendorf, P.~J. Sch{\"{u}}ffler, D.~Grolimund,
  J.~M. Buhmann, S.~Brandt, Z.~Varga, P.~J. Wild, D.~G{\"{u}}nther, and
  B.~Bodenmiller}.
\newblock \href{https://doi.org/10.1038/nmeth.2869}{Highly multiplexed imaging
  of tumor tissues with subcellular resolution by mass cytometry}.
\newblock \href{https://doi.org/10.1038/nmeth.2869}{{\em Nature Methods}},
  \href{https://doi.org/10.1038/nmeth.2869}{11(4):417--422},
  \href{https://doi.org/10.1038/nmeth.2869}{2014}.
  \href{https://doi.org/10.1038/nmeth.2869}
{doi: {{%
10\hspace{.1pt}\discretionary{.}{%
}{.}\hspace{.4pt}1038\discretionary{/}{%
}{/}nmeth\hspace{.1pt}\discretionary{.}{%
}{.}\hspace{.4pt}2869}}}


\bibitem{Gleicher2011}
\href{https://doi.org/10.1177/1473871611416549}{M.~Gleicher, D.~Albers,
  R.~Walker, I.~Jusufi, C.~D. Hansen, and J.~C. Roberts}.
\newblock \href{https://doi.org/10.1177/1473871611416549}{Visual comparison for
  information visualization}.
\newblock \href{https://doi.org/10.1177/1473871611416549}{{\em Information
  Visualization}}, \href{https://doi.org/10.1177/1473871611416549}{10(4)},
  \href{https://doi.org/10.1177/1473871611416549}{2011}.
  \href{https://doi.org/10.1177/1473871611416549}
{doi: {{%
10\hspace{.1pt}\discretionary{.}{%
}{.}\hspace{.4pt}1177\discretionary{/}{%
}{/}1473871611416549}}}


\bibitem{Goltsev2018}
\href{https://doi.org/10.1016/j.cell.2018.07.010}{Y.~Goltsev, N.~Samusik,
  J.~Kennedy-Darling, S.~Bhate, M.~Hale, G.~Vazquez, S.~Black, and G.~P.
  Nolan}.
\newblock \href{https://doi.org/10.1016/j.cell.2018.07.010}{Deep profiling of
  mouse splenic architecture with codex multiplexed imaging}.
\newblock \href{https://doi.org/10.1016/j.cell.2018.07.010}{{\em Cell}},
  \href{https://doi.org/10.1016/j.cell.2018.07.010}{174(4):968--981.e15},
  \href{https://doi.org/10.1016/j.cell.2018.07.010}{2018}.
  \href{https://doi.org/10.1016/j.cell.2018.07.010}
{doi: {{%
10\hspace{.1pt}\discretionary{.}{%
}{.}\hspace{.4pt}1016\discretionary{/}{%
}{/}j\hspace{.1pt}\discretionary{.}{%
}{.}\hspace{.4pt}cell\hspace{.1pt}\discretionary{.}{%
}{.}\hspace{.4pt}2018\hspace{.1pt}\discretionary{.}{%
}{.}\hspace{.4pt}07\hspace{.1pt}\discretionary{.}{%
}{.}\hspace{.4pt}010}}}


\bibitem{heer2012interactive}
\href{https://doi.org/10.1145/2133806.2133821}{J.~Heer and B.~Shneiderman}.
\newblock \href{https://doi.org/10.1145/2133806.2133821}{Interactive dynamics
  for visual analysis}.
\newblock \href{https://doi.org/10.1145/2133806.2133821}{{\em Communications of
  the ACM}}, \href{https://doi.org/10.1145/2133806.2133821}{55(4):45–54},
  \href{https://doi.org/10.1145/2133806.2133821}{2012}.
  \href{https://doi.org/10.1145/2133806.2133821}
{doi: {{%
10\hspace{.1pt}\discretionary{.}{%
}{.}\hspace{.4pt}1145\discretionary{/}{%
}{/}2133806\hspace{.1pt}\discretionary{.}{%
}{.}\hspace{.4pt}2133821}}}


\bibitem{Nghiem2016}
\href{https://doi.org/10.1002/cjp2.113}{M.~E. Ijsselsteijn, T.~P. Brouwer,
  Z.~Abdulrahman, E.~Reidy, A.~Ramalheiro, A.~M. Heeren, A.~Vahrmeijer, E.~S.
  Jordanova, and N.~F. de~Miranda}.
\newblock \href{https://doi.org/10.1002/cjp2.113}{Cancer immunophenotyping by
  seven-colour multispectral imaging without tyramide signal amplification}.
\newblock \href{https://doi.org/10.1002/cjp2.113}{{\em The Journal of
  Pathology: Clinical Research}},
  \href{https://doi.org/10.1002/cjp2.113}{5:3--11},
  \href{https://doi.org/10.1002/cjp2.113}{2019}.
  \href{https://doi.org/10.1002/cjp2.113}
{doi: {{%
10\hspace{.1pt}\discretionary{.}{%
}{.}\hspace{.4pt}1002\discretionary{/}{%
}{/}cjp2\hspace{.1pt}\discretionary{.}{%
}{.}\hspace{.4pt}113}}}


\bibitem{Jackson2020}
\href{https://doi.org/10.1038/s41586-019-1876-x}{H.~W. Jackson, J.~R. Fischer,
  V.~R. Zanotelli, H.~R. Ali, R.~Mechera, S.~D. Soysal, H.~Moch, S.~Muenst,
  Z.~Varga, W.~P. Weber, and B.~Bodenmiller}.
\newblock \href{https://doi.org/10.1038/s41586-019-1876-x}{The single-cell
  pathology landscape of breast cancer}.
\newblock \href{https://doi.org/10.1038/s41586-019-1876-x}{{\em Nature}},
  \href{https://doi.org/10.1038/s41586-019-1876-x}{578(7796):615--620},
  \href{https://doi.org/10.1038/s41586-019-1876-x}{2020}.
  \href{https://doi.org/10.1038/s41586-019-1876-x}
{doi: {{%
10\hspace{.1pt}\discretionary{.}{%
}{.}\hspace{.4pt}1038\discretionary{/}{%
}{/}s41586\discretionary{%
}{-}{-}019\discretionary{%
}{-}{-}1876\discretionary{%
}{-}{-}x}}}


\bibitem{kampstra2008beanplot}
\href{https://doi.org/10.18637/jss.v028.c01}{P.~Kampstra}.
\newblock \href{https://doi.org/10.18637/jss.v028.c01}{Beanplot: A boxplot
  alternative for visual comparison of distributions}.
\newblock \href{https://doi.org/10.18637/jss.v028.c01}{{\em Journal of
  Statistical Software}},
  \href{https://doi.org/10.18637/jss.v028.c01}{28(1):1--9},
  \href{https://doi.org/10.18637/jss.v028.c01}{2008}.
  \href{https://doi.org/10.18637/jss.v028.c01}
{doi: {{%
10\hspace{.1pt}\discretionary{.}{%
}{.}\hspace{.4pt}18637\discretionary{/}{%
}{/}jss\hspace{.1pt}\discretionary{.}{%
}{.}\hspace{.4pt}v028\hspace{.1pt}\discretionary{.}{%
}{.}\hspace{.4pt}c01}}}


\bibitem{Ke2013}
\href{https://doi.org/10.1038/nmeth.2563}{R.~Ke, M.~Mignardi, A.~Pacureanu,
  J.~Svedlund, J.~Botling, C.~W{\"{a}}hlby, and M.~Nilsson}.
\newblock \href{https://doi.org/10.1038/nmeth.2563}{In situ sequencing for rna
  analysis in preserved tissue and cells}.
\newblock \href{https://doi.org/10.1038/nmeth.2563}{{\em Nature Methods}},
  \href{https://doi.org/10.1038/nmeth.2563}{10(9):857--860},
  \href{https://doi.org/10.1038/nmeth.2563}{2013}.
  \href{https://doi.org/10.1038/nmeth.2563}
{doi: {{%
10\hspace{.1pt}\discretionary{.}{%
}{.}\hspace{.4pt}1038\discretionary{/}{%
}{/}nmeth\hspace{.1pt}\discretionary{.}{%
}{.}\hspace{.4pt}2563}}}


\bibitem{Keren2020}
\href{https://doi.org/10.1038/s43018-020-0031-9}{L.~Keren and M.~Angelo}.
\newblock \href{https://doi.org/10.1038/s43018-020-0031-9}{Mapping cell
  phenotypes in breast cancer}.
\newblock \href{https://doi.org/10.1038/s43018-020-0031-9}{{\em Nature
  Cancer}}, \href{https://doi.org/10.1038/s43018-020-0031-9}{1(2):156--157},
  \href{https://doi.org/10.1038/s43018-020-0031-9}{2020}.
  \href{https://doi.org/10.1038/s43018-020-0031-9}
{doi: {{%
10\hspace{.1pt}\discretionary{.}{%
}{.}\hspace{.4pt}1038\discretionary{/}{%
}{/}s43018\discretionary{%
}{-}{-}020\discretionary{%
}{-}{-}0031\discretionary{%
}{-}{-}9}}}


\bibitem{Keren2019}
\href{https://doi.org/10.1126/sciadv.aax5851}{L.~Keren, M.~Bosse, S.~Thompson,
  T.~Risom, K.~Vijayaragavan, E.~McCaffrey, D.~Marquez, R.~Angoshtari, N.~F.
  Greenwald, H.~Fienberg, J.~Wang, N.~Kambham, D.~Kirkwood, G.~Nolan, T.~J.
  Montine, S.~J. Galli, R.~West, S.~C. Bendall, and M.~Angelo}.
\newblock \href{https://doi.org/10.1126/sciadv.aax5851}{{MIBI-TOF:} a
  multiplexed imaging platform relates cellular phenotypes and tissue
  structure}.
\newblock \href{https://doi.org/10.1126/sciadv.aax5851}{{\em Science
  Advances}}, \href{https://doi.org/10.1126/sciadv.aax5851}{5(10):eaax5851},
  \href{https://doi.org/10.1126/sciadv.aax5851}{2019}.
  \href{https://doi.org/10.1126/sciadv.aax5851}
{doi: {{%
10\hspace{.1pt}\discretionary{.}{%
}{.}\hspace{.4pt}1126\discretionary{/}{%
}{/}sciadv\hspace{.1pt}\discretionary{.}{%
}{.}\hspace{.4pt}aax5851}}}


\bibitem{Krueger2020}
\href{https://doi.org/10.1109/TVCG.2019.2934547}{R.~Krueger, J.~Beyer, W.~D.
  Jang, N.~W. Kim, A.~Sokolov, P.~K. Sorger, and H.~Pfister}.
\newblock \href{https://doi.org/10.1109/TVCG.2019.2934547}{Facetto: Combining
  unsupervised and supervised learning for hierarchical phenotype analysis in
  multi-channel image data}.
\newblock \href{https://doi.org/10.1109/TVCG.2019.2934547}{{\em IEEE
  Transactions on Visualization and Computer Graphics}},
  \href{https://doi.org/10.1109/TVCG.2019.2934547}{26(1):227--237},
  \href{https://doi.org/10.1109/TVCG.2019.2934547}{2020}.
  \href{https://doi.org/10.1109/TVCG.2019.2934547}
{doi: {{%
10\hspace{.1pt}\discretionary{.}{%
}{.}\hspace{.4pt}1109\discretionary{/}{%
}{/}TVCG\hspace{.1pt}\discretionary{.}{%
}{.}\hspace{.4pt}2019\hspace{.1pt}\discretionary{.}{%
}{.}\hspace{.4pt}2934547}}}


\bibitem{Lee2015}
\href{https://doi.org/10.1038/nprot.2014.191}{J.~H. Lee, E.~R. Daugharthy,
  J.~Scheiman, R.~Kalhor, T.~C. Ferrante, R.~Terry, B.~M. Turczyk, J.~L. Yang,
  H.~S. Lee, J.~Aach, K.~Zhang, and G.~M. Church}.
\newblock \href{https://doi.org/10.1038/nprot.2014.191}{Fluorescent in situ
  sequencing {(FISSEQ)} of rna for gene expression profiling in intact cells
  and tissues}.
\newblock \href{https://doi.org/10.1038/nprot.2014.191}{{\em Nature
  Protocols}}, \href{https://doi.org/10.1038/nprot.2014.191}{10(3):442--458},
  \href{https://doi.org/10.1038/nprot.2014.191}{2015}.
  \href{https://doi.org/10.1038/nprot.2014.191}
{doi: {{%
10\hspace{.1pt}\discretionary{.}{%
}{.}\hspace{.4pt}1038\discretionary{/}{%
}{/}nprot\hspace{.1pt}\discretionary{.}{%
}{.}\hspace{.4pt}2014\hspace{.1pt}\discretionary{.}{%
}{.}\hspace{.4pt}191}}}


\bibitem{Bronstein}
\href{https://doi.org/10.2312/PE.VMV.VMV13.105-112}{F.~Lindemann, K.~Laukamp,
  A.~H. Jacobs, and K.~Hinrichs}.
\newblock \href{https://doi.org/10.2312/PE.VMV.VMV13.105-112}{Interactive
  comparative visualization of multimodal brain tumor segmentation data}.
\newblock \href{https://doi.org/10.2312/PE.VMV.VMV13.105-112}{In {\em
  Proceedings of Vision, Modeling \& Visualization (VMV)}},
  \href{https://doi.org/10.2312/PE.VMV.VMV13.105-112}{2013}.
  \href{https://doi.org/10.2312/PE.VMV.VMV13.105-112}
{doi: {{%
10\hspace{.1pt}\discretionary{.}{%
}{.}\hspace{.4pt}2312\discretionary{/}{%
}{/}PE\hspace{.1pt}\discretionary{.}{%
}{.}\hspace{.4pt}VMV\hspace{.1pt}\discretionary{.}{%
}{.}\hspace{.4pt}VMV13\hspace{.1pt}\discretionary{.}{%
}{.}\hspace{.4pt}105\discretionary{%
}{-}{-}112}}}


\bibitem{Luk2018}
\href{https://doi.org/10.1080/2162402X.2018.1507600}{S.~J. Luk, D.~M. der
  Steen, R.~S. Hagedoorn, E.~S. Jordanova, M.~W. Schilham, J.~V. Bov{\'{e}}e,
  A.~H. Cleven, J.~F. Falkenburg, K.~Szuhai, and M.~H. Heemskerk}.
\newblock \href{https://doi.org/10.1080/2162402X.2018.1507600}{{PRAME} and {HLA
  Class I} expression patterns make synovial sarcoma a suitable target for
  {PRAME} specific t-cell receptor gene therapy}.
\newblock \href{https://doi.org/10.1080/2162402X.2018.1507600}{{\em
  OncoImmunology}},
  \href{https://doi.org/10.1080/2162402X.2018.1507600}{7(12):e1507600},
  \href{https://doi.org/10.1080/2162402X.2018.1507600}{2018}.
  \href{https://doi.org/10.1080/2162402X.2018.1507600}
{doi: {{%
10\hspace{.1pt}\discretionary{.}{%
}{.}\hspace{.4pt}1080\discretionary{/}{%
}{/}2162402X\hspace{.1pt}\discretionary{.}{%
}{.}\hspace{.4pt}2018\hspace{.1pt}\discretionary{.}{%
}{.}\hspace{.4pt}1507600}}}


\bibitem{Ma2017}
\href{https://doi.org/10.2352/J.ImagingSci.Technol.2017.61.6.000000}{C.~Ma,
  F.~Pellolio, D.~A. Llano, K.~A. Stebbings, R.~V. Kenyon, and G.~E. Marai}.
\newblock
  \href{https://doi.org/10.2352/J.ImagingSci.Technol.2017.61.6.000000}{{RemBrain:}
  exploring dynamic biospatial networks with mosaic matrices and mirror
  glyphs}.
\newblock
  \href{https://doi.org/10.2352/J.ImagingSci.Technol.2017.61.6.000000}{{\em
  Journal of Imaging Science and Technology R}},
  \href{https://doi.org/10.2352/J.ImagingSci.Technol.2017.61.6.000000}{61(6):0--1},
  \href{https://doi.org/10.2352/J.ImagingSci.Technol.2017.61.6.000000}{2017}.
  \href{https://doi.org/10.2352/J.ImagingSci.Technol.2017.61.6.000000}
{doi: {{%
10\hspace{.1pt}\discretionary{.}{%
}{.}\hspace{.4pt}2352\discretionary{/}{%
}{/}J\hspace{.1pt}\discretionary{.}{%
}{.}\hspace{.4pt}ImagingSci\hspace{.1pt}\discretionary{.}{%
}{.}\hspace{.4pt}Technol\hspace{.1pt}\discretionary{.}{%
}{.}\hspace{.4pt}2017\hspace{.1pt}\discretionary{.}{%
}{.}\hspace{.4pt}61\hspace{.1pt}\discretionary{.}{%
}{.}\hspace{.4pt}6\hspace{.1pt}\discretionary{.}{%
}{.}\hspace{.4pt}000000}}}


\bibitem{Maries2013}
\href{https://doi.org/10.1109/TVCG.2013.161}{A.~Maries, N.~Mays, M.~Hunt, K.~F.
  Wong, W.~Layton, R.~Boudreau, C.~Rosano, and G.~E. Marai}.
\newblock \href{https://doi.org/10.1109/TVCG.2013.161}{{GRACE:} a visual
  comparison framework for integrated spatial and non-spatial geriatric data}.
\newblock \href{https://doi.org/10.1109/TVCG.2013.161}{{\em IEEE Transactions
  on Visualization and Computer Graphics}},
  \href{https://doi.org/10.1109/TVCG.2013.161}{19(12):2916--2925},
  \href{https://doi.org/10.1109/TVCG.2013.161}{2013}.
  \href{https://doi.org/10.1109/TVCG.2013.161}
{doi: {{%
10\hspace{.1pt}\discretionary{.}{%
}{.}\hspace{.4pt}1109\discretionary{/}{%
}{/}TVCG\hspace{.1pt}\discretionary{.}{%
}{.}\hspace{.4pt}2013\hspace{.1pt}\discretionary{.}{%
}{.}\hspace{.4pt}161}}}


\bibitem{Mayeux2004}
\href{https://doi.org/10.1602/neurorx.1.2.182}{R.~Mayeux}.
\newblock \href{https://doi.org/10.1602/neurorx.1.2.182}{Biomarkers: Potential
  uses and limitations}.
\newblock \href{https://doi.org/10.1602/neurorx.1.2.182}{{\em NeuroRX}},
  \href{https://doi.org/10.1602/neurorx.1.2.182}{1(2):182--188},
  \href{https://doi.org/10.1602/neurorx.1.2.182}{2004}.
  \href{https://doi.org/10.1602/neurorx.1.2.182}
{doi: {{%
10\hspace{.1pt}\discretionary{.}{%
}{.}\hspace{.4pt}1602\discretionary{/}{%
}{/}neurorx\hspace{.1pt}\discretionary{.}{%
}{.}\hspace{.4pt}1\hspace{.1pt}\discretionary{.}{%
}{.}\hspace{.4pt}2\hspace{.1pt}\discretionary{.}{%
}{.}\hspace{.4pt}182}}}


\bibitem{Nagaishi2011}
\href{https://doi.org/10.1136/jnnp-2011-300403}{A.~Nagaishi, M.~Takagi,
  A.~Umemura, M.~Tanaka, Y.~Kitagawa, M.~Matsui, M.~Nishizawa, K.~Sakimura, and
  K.~Tanaka}.
\newblock \href{https://doi.org/10.1136/jnnp-2011-300403}{Clinical features of
  neuromyelitis optica in a large japanese cohort: Comparison between
  phenotypes}.
\newblock \href{https://doi.org/10.1136/jnnp-2011-300403}{{\em Journal of
  Neurology, Neurosurgery and Psychiatry}},
  \href{https://doi.org/10.1136/jnnp-2011-300403}{82(12):1360--1364},
  \href{https://doi.org/10.1136/jnnp-2011-300403}{2011}.
  \href{https://doi.org/10.1136/jnnp-2011-300403}
{doi: {{%
10\hspace{.1pt}\discretionary{.}{%
}{.}\hspace{.4pt}1136\discretionary{/}{%
}{/}jnnp\discretionary{%
}{-}{-}2011\discretionary{%
}{-}{-}300403}}}


\bibitem{newschaffer2000causes}
\href{https://doi.org/10.1093/jnci/92.8.613}{C.~J. Newschaffer, K.~Otani, M.~K.
  McDonald, and L.~T. Penberthy}.
\newblock \href{https://doi.org/10.1093/jnci/92.8.613}{Causes of death in
  elderly prostate cancer patients and in a comparison nonprostate cancer
  cohort}.
\newblock \href{https://doi.org/10.1093/jnci/92.8.613}{{\em Journal of the
  National Cancer Institute}},
  \href{https://doi.org/10.1093/jnci/92.8.613}{92(8):613--621},
  \href{https://doi.org/10.1093/jnci/92.8.613}{2000}.
  \href{https://doi.org/10.1093/jnci/92.8.613}
{doi: {{%
10\hspace{.1pt}\discretionary{.}{%
}{.}\hspace{.4pt}1093\discretionary{/}{%
}{/}jnci\discretionary{/}{%
}{/}92\hspace{.1pt}\discretionary{.}{%
}{.}\hspace{.4pt}8\hspace{.1pt}\discretionary{.}{%
}{.}\hspace{.4pt}613}}}


\bibitem{pagendarm1995comparative}
H.-G. Pagendarm and F.~H. Post.
\newblock Comparative visualization: Approaches and examples.
\newblock In M.~Göbel, H.~Müller, and B.~Urban, eds., {\em Visualization in
  Scientific Computing}. Springer, 1995.

\bibitem{Preim2016}
\href{https://doi.org/10.1007/978-3-319-24523-2_10}{B.~Preim, P.~Klemm,
  H.~Hauser, K.~Hegenscheid, S.~Oeltze, K.~Toennies, and H.~V{\"{o}}lzke}.
\newblock \href{https://doi.org/10.1007/978-3-319-24523-2_10}{Visual analytics
  of image-centric cohort studies in epidemiology}.
\newblock \href{https://doi.org/10.1007/978-3-319-24523-2_10}{In L.~Linsen,
  B.~Hamann, and H.~C. Hege, eds., {\em Visualization in Medicine and Life
  Sciences III. Mathematics and Visualization.}},
  \href{https://doi.org/10.1007/978-3-319-24523-2_10}{pp. 221--248}.
  \href{https://doi.org/10.1007/978-3-319-24523-2_10}{Springer},
  \href{https://doi.org/10.1007/978-3-319-24523-2_10}{2016}.
  \href{https://doi.org/10.1007/978-3-319-24523-2_10}
{doi: {{%
10\hspace{.1pt}\discretionary{.}{%
}{.}\hspace{.4pt}1007\discretionary{/}{%
}{/}978\discretionary{%
}{-}{-}3\discretionary{%
}{-}{-}319\discretionary{%
}{-}{-}24523\discretionary{%
}{-}{-}2\_10}}}


\bibitem{Raidou2018}
\href{https://doi.org/10.1111/cgf.13413}{R.~Raidou, O.~Casares-Magaz,
  A.~Amirkhanov, V.~Moiseenko, L.~Muren, J.~Einck, A.~Vilanova, and
  M.~Gr{\"{o}}ller}.
\newblock \href{https://doi.org/10.1111/cgf.13413}{{Bladder Runner} : Visual
  analytics for the exploration of rt-induced bladder toxicity in a cohort
  study}.
\newblock \href{https://doi.org/10.1111/cgf.13413}{{\em Computer Graphics
  Forum}}, \href{https://doi.org/10.1111/cgf.13413}{37(3):205--216},
  \href{https://doi.org/10.1111/cgf.13413}{2018}.
  \href{https://doi.org/10.1111/cgf.13413}
{doi: {{%
10\hspace{.1pt}\discretionary{.}{%
}{.}\hspace{.4pt}1111\discretionary{/}{%
}{/}cgf\hspace{.1pt}\discretionary{.}{%
}{.}\hspace{.4pt}13413}}}


\bibitem{Rashid2020.03.27.001834}
\href{https://doi.org/10.1101/2020.03.27.001834}{R.~Rashid, Y.-A. Chen,
  J.~Hoffer, J.~L. Muhlich, J.-R. Lin, R.~Krueger, H.~Pfister, R.~Mitchell,
  S.~Santagata, and P.~K. Sorger}.
\newblock \href{https://doi.org/10.1101/2020.03.27.001834}{Interpretative
  guides for interacting with tissue atlas and digital pathology data using the
  minerva browser}.
\newblock \href{https://doi.org/10.1101/2020.03.27.001834}{{\em bioRxiv}},
  \href{https://doi.org/10.1101/2020.03.27.001834}{2020}.
  \href{https://doi.org/10.1101/2020.03.27.001834}
{doi: {{%
10\hspace{.1pt}\discretionary{.}{%
}{.}\hspace{.4pt}1101\discretionary{/}{%
}{/}2020\hspace{.1pt}\discretionary{.}{%
}{.}\hspace{.4pt}03\hspace{.1pt}\discretionary{.}{%
}{.}\hspace{.4pt}27\hspace{.1pt}\discretionary{.}{%
}{.}\hspace{.4pt}001834}}}


\bibitem{Robert2014}
\href{https://doi.org/10.1016/S0140-6736(14)60958-2}{C.~Robert, A.~Ribas, J.~D.
  Wolchok, F.~S. Hodi, O.~Hamid, R.~Kefford, J.~S. Weber, A.~M. Joshua, W.-J.
  Hwu, T.~C. Gangadhar, A.~Patnaik, R.~Dronca, H.~Zarour, R.~W. Joseph,
  P.~Boasberg, B.~Chmielowski, C.~Mateus, M.~A. Postow, K.~Gergich,
  J.~Elassaiss-Schaap, X.~N. Li, R.~Iannone, S.~W. Ebbinghaus, S.~P. Kang, and
  A.~Daud}.
\newblock
  \href{https://doi.org/10.1016/S0140-6736(14)60958-2}{Anti-programmed-death-receptor-1
  treatment with pembrolizumab in ipilimumab-refractory advanced melanoma: A
  randomised dose-comparison cohort of a phase 1 trial}.
\newblock \href{https://doi.org/10.1016/S0140-6736(14)60958-2}{{\em The
  Lancet}},
  \href{https://doi.org/10.1016/S0140-6736(14)60958-2}{384(9948):1109--1117},
  \href{https://doi.org/10.1016/S0140-6736(14)60958-2}{2014}.
  \href{https://doi.org/10.1016/S0140-6736(14)60958-2}
{doi: {{%
10\hspace{.1pt}\discretionary{.}{%
}{.}\hspace{.4pt}1016\discretionary{/}{%
}{/}S0140\discretionary{%
}{-}{-}6736\discretionary{%
}{(}{(}14\discretionary{)}{%
}{)}60958\discretionary{%
}{-}{-}2}}}


\bibitem{Rousseeuw1987}
\href{https://doi.org/10.1016/0377-0427(87)90125-7}{P.~J. Rousseeuw}.
\newblock \href{https://doi.org/10.1016/0377-0427(87)90125-7}{Silhouettes: A
  graphical aid to the interpretation and validation of cluster analysis}.
\newblock \href{https://doi.org/10.1016/0377-0427(87)90125-7}{{\em Journal of
  Computational and Applied Mathematics}},
  \href{https://doi.org/10.1016/0377-0427(87)90125-7}{20(C):53--65},
  \href{https://doi.org/10.1016/0377-0427(87)90125-7}{1987}.
  \href{https://doi.org/10.1016/0377-0427(87)90125-7}
{doi: {{%
10\hspace{.1pt}\discretionary{.}{%
}{.}\hspace{.4pt}1016\discretionary{/}{%
}{/}0377\discretionary{%
}{-}{-}0427\discretionary{%
}{(}{(}87\discretionary{)}{%
}{)}90125\discretionary{%
}{-}{-}7}}}


\bibitem{Schapiro2017}
\href{https://doi.org/10.1038/nmeth.4391}{D.~Schapiro, H.~W. Jackson,
  S.~Raghuraman, J.~R. Fischer, V.~R. Zanotelli, D.~Schulz, C.~Giesen,
  R.~Catena, Z.~Varga, and B.~Bodenmiller}.
\newblock \href{https://doi.org/10.1038/nmeth.4391}{{HistoCAT:} analysis of
  cell phenotypes and interactions in multiplex image cytometry data}.
\newblock \href{https://doi.org/10.1038/nmeth.4391}{{\em Nature Methods}},
  \href{https://doi.org/10.1038/nmeth.4391}{14(9):873--876},
  \href{https://doi.org/10.1038/nmeth.4391}{2017}.
  \href{https://doi.org/10.1038/nmeth.4391}
{doi: {{%
10\hspace{.1pt}\discretionary{.}{%
}{.}\hspace{.4pt}1038\discretionary{/}{%
}{/}nmeth\hspace{.1pt}\discretionary{.}{%
}{.}\hspace{.4pt}4391}}}


\bibitem{Schmidt2013}
\href{https://doi.org/10.1109/TVCG.2013.213}{J.~Schmidt, M.~E. Gr\"oller, and
  S.~Bruckner}.
\newblock \href{https://doi.org/10.1109/TVCG.2013.213}{{VAICo:} visual analysis
  for image comparison}.
\newblock \href{https://doi.org/10.1109/TVCG.2013.213}{{\em IEEE Transactions
  on Visualization and Computer Graphics}},
  \href{https://doi.org/10.1109/TVCG.2013.213}{19(12):2090--2099},
  \href{https://doi.org/10.1109/TVCG.2013.213}{2013}.
  \href{https://doi.org/10.1109/TVCG.2013.213}
{doi: {{%
10\hspace{.1pt}\discretionary{.}{%
}{.}\hspace{.4pt}1109\discretionary{/}{%
}{/}TVCG\hspace{.1pt}\discretionary{.}{%
}{.}\hspace{.4pt}2013\hspace{.1pt}\discretionary{.}{%
}{.}\hspace{.4pt}213}}}


\bibitem{schuffler2015automatic}
\href{https://doi.org/10.1186/s12859-014-0431-x}{P.~J. Sch{\"u}ffler,
  D.~Schapiro, C.~Giesen, H.~A. Wang, B.~Bodenmiller, and J.~M. Buhmann}.
\newblock \href{https://doi.org/10.1186/s12859-014-0431-x}{Automatic single
  cell segmentation on highly multiplexed tissue images}.
\newblock \href{https://doi.org/10.1186/s12859-014-0431-x}{{\em Cytometry Part
  A}}, \href{https://doi.org/10.1186/s12859-014-0431-x}{87(10):936--942},
  \href{https://doi.org/10.1186/s12859-014-0431-x}{2015}.
  \href{https://doi.org/10.1186/s12859-014-0431-x}
{doi: {{%
10\hspace{.1pt}\discretionary{.}{%
}{.}\hspace{.4pt}1186\discretionary{/}{%
}{/}s12859\discretionary{%
}{-}{-}014\discretionary{%
}{-}{-}0431\discretionary{%
}{-}{-}x}}}


\bibitem{Shneiderman1994}
\href{https://doi.org/10.1109/52.329404}{B.~Shneiderman}.
\newblock \href{https://doi.org/10.1109/52.329404}{Dynamic queries for visual
  information seeking}.
\newblock \href{https://doi.org/10.1109/52.329404}{{\em IEEE Software}},
  \href{https://doi.org/10.1109/52.329404}{11(6):70--77},
  \href{https://doi.org/10.1109/52.329404}{1994}.
  \href{https://doi.org/10.1109/52.329404}
{doi: {{%
10\hspace{.1pt}\discretionary{.}{%
}{.}\hspace{.4pt}1109\discretionary{/}{%
}{/}52\hspace{.1pt}\discretionary{.}{%
}{.}\hspace{.4pt}329404}}}


\bibitem{Somarakis2019}
\href{https://doi.org/10.1109/tvcg.2019.2931299}{A.~Somarakis, V.~{Van Unen},
  F.~Koning, B.~P. Lelieveldt, and T.~H\"ollt}.
\newblock \href{https://doi.org/10.1109/tvcg.2019.2931299}{{ImaCytE:} visual
  exploration of cellular microenvironments for imaging mass cytometry data}.
\newblock \href{https://doi.org/10.1109/tvcg.2019.2931299}{{\em IEEE
  Transactions on Visualization and Computer Graphics}},
  \href{https://doi.org/10.1109/tvcg.2019.2931299}{pp. 1--1},
  \href{https://doi.org/10.1109/tvcg.2019.2931299}{2019}.
  \href{https://doi.org/10.1109/tvcg.2019.2931299}
{doi: {{%
10\hspace{.1pt}\discretionary{.}{%
}{.}\hspace{.4pt}1109\discretionary{/}{%
}{/}tvcg\hspace{.1pt}\discretionary{.}{%
}{.}\hspace{.4pt}2019\hspace{.1pt}\discretionary{.}{%
}{.}\hspace{.4pt}2931299}}}


\bibitem{Steenwijk2010}
M.~D. Steenwijk, J.~Milles, M.~Buchem, J.~Reiber, and C.~P. Botha.
\newblock Integrated visual analysis for heterogeneous datasets in cohort
  studies.
\newblock In {\em Proceedings of the Workshop on Visual Analytics in Healthcare
  (VAHC)}, 2010.

\bibitem{Stolte02polaris:a}
\href{https://doi.org/10.1109/INFVIS.2000.885086}{C.~Stolte and P.~Hanrahan}.
\newblock \href{https://doi.org/10.1109/INFVIS.2000.885086}{Polaris: A system
  for query, analysis and visualization of multi-dimensional relational
  databases}.
\newblock \href{https://doi.org/10.1109/INFVIS.2000.885086}{{\em IEEE
  Transactions on Visualization and Computer Graphics}},
  \href{https://doi.org/10.1109/INFVIS.2000.885086}{8:52--65},
  \href{https://doi.org/10.1109/INFVIS.2000.885086}{2002}.
  \href{https://doi.org/10.1109/INFVIS.2000.885086}
{doi: {{%
10\hspace{.1pt}\discretionary{.}{%
}{.}\hspace{.4pt}1109\discretionary{/}{%
}{/}INFVIS\hspace{.1pt}\discretionary{.}{%
}{.}\hspace{.4pt}2000\hspace{.1pt}\discretionary{.}{%
}{.}\hspace{.4pt}885086}}}


\bibitem{stoltzfus2020cytomap}
\href{https://doi.org/10.1016/j.celrep.2020.107523}{C.~R. Stoltzfus,
  J.~Filipek, B.~H. Gern, B.~E. Olin, J.~M. Leal, Y.~Wu, M.~R. Lyons-Cohen,
  J.~Y. Huang, C.~L. Paz-Stoltzfus, C.~R. Plumlee, et~al.}
\newblock \href{https://doi.org/10.1016/j.celrep.2020.107523}{Cytomap: A
  spatial analysis toolbox reveals features of myeloid cell organization in
  lymphoid tissues}.
\newblock \href{https://doi.org/10.1016/j.celrep.2020.107523}{{\em Cell
  reports}}, \href{https://doi.org/10.1016/j.celrep.2020.107523}{31(3):107523},
  \href{https://doi.org/10.1016/j.celrep.2020.107523}{2020}.
  \href{https://doi.org/10.1016/j.celrep.2020.107523}
{doi: {{%
10\hspace{.1pt}\discretionary{.}{%
}{.}\hspace{.4pt}1016\discretionary{/}{%
}{/}j\hspace{.1pt}\discretionary{.}{%
}{.}\hspace{.4pt}celrep\hspace{.1pt}\discretionary{.}{%
}{.}\hspace{.4pt}2020\hspace{.1pt}\discretionary{.}{%
}{.}\hspace{.4pt}107523}}}


\bibitem{tufte1990envisioning}
E.~R. Tufte.
\newblock {\em Envisioning information}.
\newblock Graphics Press, 1990.

\bibitem{VanUnen2017}
\href{https://doi.org/10.1038/s41467-017-01689-9}{V.~{Van Unen},
  T.~H{\"{o}}llt, N.~Pezzotti, N.~Li, M.~J. Reinders, E.~Eisemann, F.~Koning,
  A.~Vilanova, and B.~P. Lelieveldt}.
\newblock \href{https://doi.org/10.1038/s41467-017-01689-9}{Visual analysis of
  mass cytometry data by hierarchical stochastic neighbour embedding reveals
  rare cell types}.
\newblock \href{https://doi.org/10.1038/s41467-017-01689-9}{{\em Nature
  Communications}},
  \href{https://doi.org/10.1038/s41467-017-01689-9}{8(1):1--10},
  \href{https://doi.org/10.1038/s41467-017-01689-9}{2017}.
  \href{https://doi.org/10.1038/s41467-017-01689-9}
{doi: {{%
10\hspace{.1pt}\discretionary{.}{%
}{.}\hspace{.4pt}1038\discretionary{/}{%
}{/}s41467\discretionary{%
}{-}{-}017\discretionary{%
}{-}{-}01689\discretionary{%
}{-}{-}9}}}


\bibitem{VanUnen2016}
\href{https://doi.org/10.1016/j.immuni.2016.04.014}{V.~van Unen, N.~Li,
  I.~Molendijk, M.~Temurhan, T.~H{\"{o}}llt, A.~E. {van der Meulen-de Jong},
  H.~W. Verspaget, M.~L. Mearin, C.~J. Mulder, J.~van Bergen, B.~P. Lelieveldt,
  and F.~Koning}.
\newblock \href{https://doi.org/10.1016/j.immuni.2016.04.014}{Mass cytometry of
  the human mucosal immune system identifies tissue- and disease-associated
  immune subsets}.
\newblock \href{https://doi.org/10.1016/j.immuni.2016.04.014}{{\em Immunity}},
  \href{https://doi.org/10.1016/j.immuni.2016.04.014}{44(5):1227--1239},
  \href{https://doi.org/10.1016/j.immuni.2016.04.014}{2016}.
  \href{https://doi.org/10.1016/j.immuni.2016.04.014}
{doi: {{%
10\hspace{.1pt}\discretionary{.}{%
}{.}\hspace{.4pt}1016\discretionary{/}{%
}{/}j\hspace{.1pt}\discretionary{.}{%
}{.}\hspace{.4pt}immuni\hspace{.1pt}\discretionary{.}{%
}{.}\hspace{.4pt}2016\hspace{.1pt}\discretionary{.}{%
}{.}\hspace{.4pt}04\hspace{.1pt}\discretionary{.}{%
}{.}\hspace{.4pt}014}}}


\bibitem{Wagner2019}
\href{https://doi.org/10.1109/TVCG.2017.2785271}{M.~Wagner, D.~Slijepcevic,
  B.~Horsak, A.~Rind, M.~Zeppelzauer, and W.~Aigner}.
\newblock \href{https://doi.org/10.1109/TVCG.2017.2785271}{{KAVAGait:}
  knowledge-assisted visual analytics for clinical gait analysis}.
\newblock \href{https://doi.org/10.1109/TVCG.2017.2785271}{{\em IEEE
  Transactions on Visualization and Computer Graphics}},
  \href{https://doi.org/10.1109/TVCG.2017.2785271}{25(3):1528--1542},
  \href{https://doi.org/10.1109/TVCG.2017.2785271}{2019}.
  \href{https://doi.org/10.1109/TVCG.2017.2785271}
{doi: {{%
10\hspace{.1pt}\discretionary{.}{%
}{.}\hspace{.4pt}1109\discretionary{/}{%
}{/}TVCG\hspace{.1pt}\discretionary{.}{%
}{.}\hspace{.4pt}2017\hspace{.1pt}\discretionary{.}{%
}{.}\hspace{.4pt}2785271}}}


\bibitem{Yuan2012}
\href{https://doi.org/10.1126/scitranslmed.3004330}{Y.~Yuan, H.~Failmezger,
  O.~M. Rueda, H.~{Raza Ali}, S.~Gr{\"{a}}f, S.~F. Chin, R.~F. Schwarz,
  C.~Curtis, M.~J. Dunning, H.~Bardwell, N.~Johnson, S.~Doyle, G.~Turashvili,
  E.~Provenzano, S.~Aparicio, C.~Caldas, and F.~Markowetz}.
\newblock \href{https://doi.org/10.1126/scitranslmed.3004330}{Quantitative
  image analysis of cellular heterogeneity in breast tumors complements genomic
  profiling}.
\newblock \href{https://doi.org/10.1126/scitranslmed.3004330}{{\em Science
  Translational Medicine}},
  \href{https://doi.org/10.1126/scitranslmed.3004330}{4(157):157ra143--157ra143},
  \href{https://doi.org/10.1126/scitranslmed.3004330}{2012}.
  \href{https://doi.org/10.1126/scitranslmed.3004330}
{doi: {{%
10\hspace{.1pt}\discretionary{.}{%
}{.}\hspace{.4pt}1126\discretionary{/}{%
}{/}scitranslmed\hspace{.1pt}\discretionary{.}{%
}{.}\hspace{.4pt}3004330}}}


\bibitem{zhang2017comparative}
\href{https://doi.org/10.2312/vcbm.20171237}{C.~Zhang, T.~H{\"{o}}llt, M.~W.~A.
  Caan, E.~Eisemann, and A.~Vilanova}.
\newblock \href{https://doi.org/10.2312/vcbm.20171237}{Comparative
  visualization for diffusion tensor imaging group study at multiple levels of
  detail}.
\newblock \href{https://doi.org/10.2312/vcbm.20171237}{In {\em Proceedings of
  Visual Computing for Biology and Medicine (VCBM)}},
  \href{https://doi.org/10.2312/vcbm.20171237}{pp. 53--62},
  \href{https://doi.org/10.2312/vcbm.20171237}{2017}.
  \href{https://doi.org/10.2312/vcbm.20171237}
{doi: {{%
10\hspace{.1pt}\discretionary{.}{%
}{.}\hspace{.4pt}2312\discretionary{/}{%
}{/}vcbm\hspace{.1pt}\discretionary{.}{%
}{.}\hspace{.4pt}20171237}}}


\bibitem{Zhang}
Z.~Zhang, D.~Gotz, and A.~Perer.
\newblock Interactive visual patient cohort analysis.
\newblock In {\em Proceedings of the Workshop on Visual Analytics in Healthcare
  (VAHC)}, 2012.

\end{thebibliography}
\end{document}